# A latent variable model for survival time prediction with censoring and diverse covariates

Shannon R. McCurdy *
California Institute for Quantitative Biosciences,
University of California, Berkeley
and
Annette Molinaro
Departments of Neurological Surgery, Epidemiology, and Biostatistics,
University of California, San Francisco
and
Lior Pachter
Division of Biology and Biological Engineering and
Department of Computing and Mathematical Sciences,
California Institute of Technology

June 20, 2017

## Abstract

Fulfilling the promise of precision medicine requires accurately and precisely classifying disease states. For cancer, this includes prediction of survival time from a surfeit of covariates. Such data presents an opportunity for improved prediction, but also a challenge due to high dimensionality. Furthermore, disease populations can be heterogeneous. Integrative modeling is sensible, as the underlying hypothesis is that joint analysis of multiple covariates provides greater explanatory power than separate analyses. We propose an integrative latent variable model that combines factor analysis for various data types and an exponential Cox proportional hazards model for continuous survival time with informative censoring. The factor and Cox models are connected through low-dimensional latent variables that can be interpreted and visualized to identify subpopulations. We use this model to predict survival time. We demonstrate this model's utility in simulation and on four Cancer Genome Atlas datasets: diffuse lower-grade glioma, glioblastoma multiforme, lung adenocarcinoma, and lung squamous cell carcinoma. These datasets have small sample sizes, high-dimensional diverse covariates, and high censorship rates. We compare the predictions from our model to two alternative models. Our model outperforms in simulation and is competitive on real datasets. Furthermore, the low-dimensional visualization for diffuse lower-grade glioma displays known subpopulations.

*Shannon R. McCurdy is a Postdoctoral Fellow, UC Berkeley, Berkeley, CA 94720 (email: smccurdy@berkeley.edu); Annette Molinaro is an Associate Professor in Residence, University of California, San Francisco, San Francisco, CA 94143 (email: Annette.Molinaro@ucsf.edu); Lior Pachter is a Professor, California Institute of Technology, Pasadena, CA 91125 (email: lpachter@caltech.edu). Research reported in this publication was supported by the National Human Genome Research Institute of the National Institutes of Health under Award Number F32HG008713. The content is solely the responsibility of the authors and does not necessarily represent the official views of the National Institutes of Health. SRM would like to thank the Simons Institute for the Theory of Computing for support, and Faraz Tavakoli, Isaac Joseph, Robert Tunney, Lenka Matejovicova, Harold Pimentel, Ilan Shomorony, Elaine Angelino, and Shishi Luo for useful comments.

# 1. INTRODUCTION

In a typical observational survival analysis, a medical study collects patient data from a cohort of individuals with a particular medical condition, usually numbering in the hundreds or thousands of individuals. During the study, the time of an event of interest, such as a recurrence or death, is recorded for each individual. Sometimes, an individual drops out of the study, or the study ends before the event of interest occurs. This is called censoring, and whether censoring occurs is also recorded for each individual, i.e. the last time the patient was seen event-free. The remainder of patient data collected can comprise genomic, epigenetic, clinical, histopathological, and demographic variables. With the inclusion of genomic and epigenetic data, the number of covariates collected for each individual can easily exceed tens of thousands. The goal of the survival analysis is to identify meaningful covariates associated with the event time.

The instantaneous risk of the event of interest given the patient data is called the hazard function. We will focus on a simple, widely-used model for the hazard function: the exponential Cox proportional hazards (ECPH) model, also called exponential regression (see Kalbfleisch and Prentice 2002). This model is contained in a wider class of models called relative risk or Cox models (Cox 1972). For Cox models, the hazard function is decomposed into the product of two functions. One function, the baseline hazards function, depends only on the event time. The other function, the relative risk function, can depend both on the event time and the measured covariates. The relative risk function is typically given as an exponential function of a linear combination of covariates. Cox proportional hazards models (CPH) have additional restrictions on the relative risk function; for these models, the covariates are time-independent. For the CPH model, the ratio of hazards for two individuals is time-independent and depends only on the difference of their covariates. There are two approaches to modeling the baseline hazards function: parametric and non-parametric. In the parametric approach, the baseline hazards function is given a specific functional form. In the non-parametric approach, the functional form of the baseline hazards function is unspecified. The ECPH model takes the parametric approach with a time-independent baseline hazards function. Here, we will focus on parametric approaches.

The ECPH model has a simple extension to include non-informative censoring (ECPH-

C). In the censored setting, there are two potential outcomes: the event of interest, and the censoring event; only the event that occurs first is observed. One can take the model for each event to be an ECPH model. The likelihood of the observed data is the cumulative incidence function for each event. Censoring is considered to be non-informative if the censoring time is conditionally independent of the event of interest given the covariates. This condition is also known as coarsening at random (CAR) (Gill *et al.* 1997). Censoring is informative if the coarsening at random condition does not hold, and fully specified parametric models are necessary (Scharfstein and Robins 2002).

The ECPH-C model requires modification in the setting where the number of covariates of the patient data approaches or exceeds the number of patients, i.e. in the high-dimensional setting. Analogous to high-dimensional linear regression, it becomes necessary to regularize the likelihood of the data when the number of covariates is large. In addition, not all of the patient covariates may be predictive of the event of interest, and a model that incorporates sparsity in the parameters is desirable. One standard approach, introduced by Tibshirani (1997), is to include the sparsity-inducing $L_1$-penalty in the likelihood. We will call the $L_1$ penalized ECPH-C model ECPH-C-$L_1$. Other high-dimensional variable selection methods in CPH models are considered in Fan *et al.* (2010).

ECPH-C-$L_1$ models and the non-parametric counterparts are simple, widely used models. Parameter estimation requires only standard iterative techniques, and these models have reasonable performance for many problems. These models assume that linear combinations of measured variables are directly responsible for the event of interest, that there is no error in the measurement, and that the measured covariates have colinearity Larsen (2004). Latent variable models were introduced to relax these stringent assumptions. Larsen (2004) considers a continuous-time latent variable model with a categorical latent space, fixed and binary covariates, and non-informative censoring. Muthn and Masyn (2005) consider a discrete-time, mixed categorical and multivariate continuous latent space model with fixed, binary, categorical, and continuous covariates, non-informative censoring, and with missing data estimation. Larsen (2005) considers a continuous-time, univariate continuous latent space model with fixed and binary covariates, and non-informative censoring. Asparouhov and Masyn (2006) consider a continuous-time, mixed categorical and univariate continuous latent space model with fixed, binary, categorical, and continuous covariates, as well as non-informative censoring. Recently, Wong *et al.* (2017) extended structural equation modeling, a continuous latent variable framework with fixed and parametrically distributed covariates, to include non-informative censoring.

In the non-survival setting, there has been recent initiative towards integrative modeling for clinical data (Kristensen *et al.* 2014 and references). The motivation behind integrative modeling is that the disparate genomic and clinical datatypes all capture different elements of the biological disease mechanisms, and that a comprehensive model will have greater predictive power than separate models for each datatype.

One type of integrative model is factor analysis, which falls within the wider class of latent variable models. Factor analysis has a long history (see Bartholomew *et al.* 2011 for detailed development), and it has been successfully put to use in clinical research to identify different subtypes of breast, lung, brain, and colorectal cancer (Shen *et al.* 2009, 2012; Mo *et al.* 2013). The central modeling hypothesis of factor analysis is that a low-dimensional, continuous latent variable is responsible for the observed covariance between the high-dimensional covariates. The observed covariates for each individual are a noisy linear combination of that individual's realization of the latent variable; in other words, the measured variables each capture certain aspects of the true variables. The estimation of each individual's realization of the latent variable given their observed data provides a low-dimensional projection. Wedel and Kamakura (2001) consider generalizations of the distributions of the observed covariates. Subpopulations identified from the latent space of factor analysis have been used in comparative Kaplan-Meier analyses (Shen *et al.* 2009, 2012). Pepke and Ver Steeg (2017) use an alternate discrete hierarchical latent variable model to identify latent factors, which they also use to perform comparative Kaplan-Meier analyses. To our knowledge, the low-dimensional latent variables from factor analysis have not been used in integrative predictive models for continuous survival time with informative censoring, and have not had demonstrated utility for heterogeneous and non-heterogenous populations.

Here, we propose a joint factor model and an exponential Cox proportional hazards model with diverse covariates (FA-ECPH). The latent space of factor analysis serves as the covariates in the ECPH model for the event of interest. With informative censoring, the latent space of factor analysis also serves as the covariates for an ECPH model for the censoring event (FA-ECPH-C). The likelihood of the observed data is the cumulative incidence function for each event integrated over the latent space. We find that a decoupled, two-part model is an excellent fast approximation to the joint model (see Sec. SM-5 of the Supplementary Materials [SM]).

This work provides several contributions to the literature. The prior body of work on latent variable survival models has not considered the high-dimensional setting. The low-dimensional latent space provides a form of dimensional reduction akin to principle component analysis. Most of the prior work has a univariate continuous latent space and none of the prior work has commented on the value of the low-dimensional projection of the data for identifying heterogeneity in the individuals. Our central modeling hypothesis is that the large-dimensional, diverse datatypes are correlated, noisy estimates of hidden variables predictive of survival. As a result, we choose a model where all of the covariance between the covariates is mediated through the low-dimensional latent-space. This means that we do not include separate effects from fixed covariates, and, unlike the prior work, our model is fully generative. Another consequence is that our model provides a regularized estimate of the covariance between covariates, although we do not pursue this here. Our model selects the direction of the latent space most predictive of event time. Aspects of the data that are not predictive of event time but are strongly correlated in the data will be captured by the other directions of the latent space. Lastly, our model allows for informative censoring; both the survival time and the censoring time depends on the latent space.

We explore the attributes of our model in a simulation study and several real data examples. We compare the accuracy of our predictions, measured by the concordance index (c-index), to the ECPH-C-$L_1$ model. In the cancer setting, where gold-standard covariates have been identified, we also compare our predictions to the ECPH-C model with the gold-standard covariates. The FA-ECPH-C model outperforms the ECPH-C-$L_1$ alternative on simulated datasets. On real data, we find that the model performance varies strongly according to the dataset, with no clear model winner across all four datasets. Lastly, we visualize the samples of the datasets according to the low-dimensional latent projections of FA-ECPH-C. This projection is compelling for diffuse lower-grade glioma. The results are discussed in Sec. 3 and 4.

## 2. METHODS

The development of the joint factor model and an exponential Cox proportional hazards model with diverse covariates is described in three parts. Sec. 2.1 comprises the conditional survival time and censoring variables. Sec. 2.2 comprises the diverse conditional covariates. Sec. 2.3 comprises the joint latent variable model. We focus on the observational study design with n={1, ..., N} individuals. The results in the remainder of this section are for a single individual. Complete details are given in the SM.

## 2.1 Survival time and censoring—exponential Cox proportional hazards

We take the ECPH-C model for the conditional survival time and censoring time distributions. The data for this part of the model is a survival time $t$ and a censoring time $c$, conditioned on the covariates, which in our setting are the latent variables $\mathbf{z}$. The latent variables are a $d_z \times 1$ vector, and it will also be convenient to introduce a $(d_z + 1) \times 1$ vector $\tilde{\mathbf{z}} = (1, \mathbf{z}^T)^T$. The latent variables are discussed in more detail in (Sec. 2.3). The conditional distributions for $t$ and $c$ given the latent variables $\mathbf{z}$ are exponential with the following parameterization for $\rho$, the standard rate parameter of the exponential distribution,

$$t|\mathbf{z} \sim \text{Exp}(\rho_T = \exp\left(\mathbf{w}_T^T \tilde{\mathbf{z}}\right)) \qquad c|\mathbf{z} \sim \text{Exp}(\rho_C = \exp\left(\mathbf{w}_C^T \tilde{\mathbf{z}}\right)). \qquad (1)$$

For this model, the conditional hazard function is $\rho_{T,C}$ for $(t,c)$ respectively. We use the subscript $\{T,C\}$ for generic expressions for the respective $(t,c)$ distributions. In Eqn. (1),

$$\mathbf{w}_{T,C} = (\ln \lambda_{T,C}, \boldsymbol{\beta_{T,C}}^T)^T, \qquad (2)$$

where $\lambda_{T,C}$ is the time-independent baseline hazard and $\boldsymbol{\beta}_{T,C}$ is the $d_z \times 1$-dimensional vector of exponential regression parameters that estimate the exponential effect size of each component of $\mathbf{z}$ on the relative risk.

The joint conditional distribution of $t$ and $c$ is,

$$p_{\mathbf{w}_T,\mathbf{w}_C}(t,c|\mathbf{z}) = p_{\mathbf{w}_T}(t|\mathbf{z})p_{\mathbf{w}_C}(c|\mathbf{z}). \qquad (3)$$

We are not able to observe both $t$ and $c$ for individuals in a medical study. Instead, we observe only the first event time, $\tilde{t} = \min(t,c)$, and whether or not the event occured, $\delta = \mathbf{I}(\tilde{t} = t)$, where $\mathbf{I}$ is the indicator function. The conditional time-to-event and censoring distribution $p(\tilde{t}, \delta|\mathbf{z})$ for an individual is simply the cumulative probability for that individual's event,

$$p_{\mathbf{w}_T,\mathbf{w}_C}(\tilde{t}, \delta|\mathbf{z}) = \left(p_{\mathbf{w}_T}(\tilde{t}|\mathbf{z})P_{\mathbf{w}_C}(c > \tilde{t}|\mathbf{z})\right)^{\delta} \left(P_{\mathbf{w}_T}(t > \tilde{t}|\mathbf{z})p_{\mathbf{w}_C}(c = \tilde{t}|\mathbf{z})\right)^{1-\delta}, \qquad (4)$$

where $P_{\mathbf{w}_{T,C}}((t,c) > \tilde{t}|\mathbf{z})$ is the conditional survival function for $(t,c)$, respectively.

ECPH-type models are discussed in more detail in Sec. SM-2.

## 2.2 Diverse covariates—factor analysis

We take the diverse conditional distributions for the observed covariates for $D$ different datatypes from FA. For each datatype $d = \{1, \ldots, D\}$, the covariates for that datatype $\mathbf{x}^{(d)}$, a vector of dimension $d_x^{(d)} \times 1$. The full data for this part of the model are the covariates $\mathbf{x}^{(1)}, \ldots, \mathbf{x}^{(D)}$ conditioned on the latent variables $\mathbf{z}$. For example, $\mathbf{x}^{(1)}$ may be gene expression measurements, $\mathbf{x}^{(d)}$ may be single nucleotide polymorphisms that each take values $x_i^{(d)} \in \{0, 1, 2\}$, and $\mathbf{x}^{(D)}$ may be a categorical variable capturing ancestral origin, such that $\sum_{i=1}^{d_x^{(D)}} x_i^{(D)} = 1$. In our model, the observed covariates $\mathbf{x}^{(1)}$ would be conditionally normal, while $\mathbf{x}^{(d)}$ would be conditionally binomial with $b^{(d)} = 2$, and $\mathbf{x}^{(D)}$ would be conditionally multinomial with $b^{(D)} = 1$,

$$\mathbf{x}^{(1)}|\mathbf{z} \sim \mathcal{N}(\mathbf{W}^{(1)}\mathbf{z} + \boldsymbol{\mu}^{(1)}, \boldsymbol{\Psi}^{(1)})$$
$$\ldots$$
$$\mathbf{x}^{(d)}|\mathbf{z} \sim \text{Binomial}(b^{(d)}, \mathbf{f}^{(d)} = \sigma(\mathbf{W}^{(d)}\mathbf{z} + \boldsymbol{\mu}^{(d)})).$$
$$\ldots$$
$$\mathbf{x}^{(D)}|\mathbf{z} \sim \text{Multinomial}(b^{(D)}, \mathbf{f}^{(D)} = \text{softmax}(\mathbf{W}^{(D)}\mathbf{z} + \boldsymbol{\mu}^{(D)})). \qquad (5)$$

In this expression, $\sigma(\mathbf{x})$ is the logistic function and softmax($\mathbf{x}$) a generalization of the logistic function (see Sec. SM-3.1). These two functions map the linear combination of the latent variables, $\mathbf{Wz} + \boldsymbol{\mu}$, to the mean parameters of the conditional distributions, which in these two cases are frequencies $\mathbf{f}$. For each datatype $d$, $\mathbf{W}^{(d)}$ is a $(d_x^{(d)} \times d_z)$ matrix of factor loading parameters, $\boldsymbol{\mu}^{(d)}$ a $(d_x^{(d)} \times 1)$-dimensional vector of mean parameters, and, for conditionally normal distributions, $\boldsymbol{\Psi}^{(d)} = \text{diag}(\boldsymbol{\psi})^{(d)}$ a $(d_x^{(d)} \times d_x^{(d)})$ diagonal matrix of error parameters. For conditionally multinomial distributions to have the proper number of parameters, the elements of the last row of conditionally multinomial $\mathbf{W}$'s and $\boldsymbol{\mu}$'s are equal to zero. The factor loadings $\mathbf{W}_{ij}^{(d)}$ estimate the effect size of the latent variable $z_j$ on the $x_i^{th}$ component of the observed covariate $\mathbf{x}$. The $\boldsymbol{\mu}^{(d)}$'s allow for an intercept effect, and for conditionally normal distributions, $\boldsymbol{\Psi}^{(d)}$ allows for additional heteroscedastic error. Note the modeling assumption of conditional independence, given the latent variables $\mathbf{z}$, between all $i \in \{1, \ldots, d_x^{(d)}\}$ components of the conditionally normal and conditionally binomial data $x_i^{(d)}$. The conditional independence assumption will also hold for the components of the conditionally multinomial data under a variational approximation explained in detail in Sec. SM-3.1.

The joint conditional probability of the observed covariates is,

$$p_\Theta(\mathbf{x}^{(1)}, \ldots, \mathbf{x}^{(D)} | \mathbf{z}) = \prod_{d=1}^{D} p_{\Theta^{(d)}}(\mathbf{x}^{(d)} | \mathbf{z}), \tag{6}$$

where we have collected all of the parameters for the $d^{th}$ conditional distribution into $\Theta^{(d)}$ and all of the parameters of the joint conditional distribution into $\Theta = \{\Theta^{(1)}, \ldots, \Theta^{(D)}\}$. The conditional distributions $p_{\Theta^{(d)}}(\mathbf{x}^{(d)} | \mathbf{z})$ are from Eqn. (5). Note the modeling assumption of conditional independence between each of the datatypes given the latent variables $\mathbf{z}$.

FA is discussed in more detail in Sec. SM-3.

## 2.3 Combining exponential Cox proportional hazards and factor analysis

We take the following latent variable model, FA-ECPH-C, for our observed covariates $\mathbf{x}^{(d)}$ from $D$ different datatypes and survival time with informative censoring. We have a latent variable of dimension $d_z \times 1$ that is normally distributed,

$$\mathbf{z} \sim \mathcal{N}(\mathbb{0}, \mathbb{1}). \tag{7}$$

We choose a continuous rather than discrete latent variable. Discrete latent variables are often used for clustering problems. We choose continuous latent variables because this framework is flexible enough to capture clustering, but does not require clustering *a priori*. Determining the dimension $d_z$ of the latent variable is a model selection problem.

The model for data generation is,

$$p_{\mathbf{w}_T, \mathbf{w}_C, \Theta}(t, c, \mathbf{x}^{(1)}, \ldots, \mathbf{x}^{(D)}, \mathbf{z}) = p_{\mathbf{w}_T, \mathbf{w}_C}(t, c|\mathbf{z}) p_{\Theta}(\mathbf{x}^{(1)}, \ldots, \mathbf{x}^{(D)}|\mathbf{z}) p(\mathbf{z}), \qquad (8)$$

where $p_{\mathbf{w}_T, \mathbf{w}_C}(t, c|\mathbf{z})$ is defined in Eqn. (1) and $p_{\Theta}(\mathbf{x}^{(1)}, \ldots, \mathbf{x}^{(D)}|\mathbf{z})$ is defined in Eqn. (6). The key assumption for this model is that, conditioned on the latent variable $\mathbf{z}$, the observed covariates $\mathbf{x}^{(d)}$, $t$, and $c$ are all mutually independent. Note that while $t \perp c|\mathbf{z}$, for generic $\mathbf{w}_C$, $t \not\perp c|\mathbf{x}^{(1)}, \ldots, \mathbf{x}^{(D)}$, and the CAR assumption does not hold. The CAR assumption does hold when $\boldsymbol{\beta}_C = \mathbb{0}$ (see Eqn. (2) for the relation between $\mathbf{w}_C$ and $\boldsymbol{\beta}_C$).

Recall that $t$ and $c$ are not jointly observed in the clinical setting; instead one observes an event time $\tilde{t}$ and censoring indicator $\delta$. The joint probability of the clinically observed time-to-event, censoring indicator, observed covariates, and latent variables is,

$$p_{\mathbf{w}_T, \mathbf{w}_C, \Theta}(\tilde{t}, \delta, \mathbf{x}^{(1)}, \ldots, \mathbf{x}^{(D)}, \mathbf{z}) = p_{\mathbf{w}_T, \mathbf{w}_C}(\tilde{t}, \delta|\mathbf{z}) p_{\Theta}(\mathbf{x}^{(1)}, \ldots, \mathbf{x}^{(D)}|\mathbf{z}) p(\mathbf{z}), \qquad (9)$$

where $p_{\mathbf{w}_T, \mathbf{w}_C}(\tilde{t}, \delta|\mathbf{z})$ is defined in Eqn. (4).

Given a set of $N$ independent observations, an approximate, generalized expectation-maximization (GEM) algorithm provides maximum likelihood estimates (MLE) for the parameter values $\hat{\mathbf{w}}_T, \hat{\mathbf{w}}_C, \hat{\Theta}$. In a GEM algorithm, one calculates conditional expectations of the latent variables given the observed covariates and estimates for the parameters. This is called the expectation (E) step. In our setting, an approximate algorithm for the E-step is required, because the marginal likelihood of the observed covariates $p_{\mathbf{w}_T, \mathbf{w}_C, \Theta}(\tilde{t}, \delta, \mathbf{x}^{(1)}, \ldots, \mathbf{x}^{(D)})$ is not analytic. We use Monte-Carlo methods to approximate the E-step. In the maximization (M) step of an EM algorithm, one maximizes the parameters with respect to the conditional expectation of the complete log likelihood, conditioned on the observed covariates. For GEM algorithms, the true maximum solution is not analytic, and approximate maximization is performed. In our setting, we use both Newton-Raphson and conditional maximization approaches. The inference problem is discussed in additional detail in Sec. SM-4.1.

We find a fast approximation to the fully integrated model FA-ECPH-C. The fast approximation is the decoupled, two-part model of first inference for factor analysis and estimation of the conditional expectation of the latent variables, and then second inference for an exponential Cox proportional hazards model with the conditional expectation of the latent variables as covariates. We show details in Sec. SM-5. This shows that the local maximum of the likelihood function picked out by the decoupled model is also a local maximum for the fully integrated model.

Once we have maximum likelihood estimates for the parameter values, we can predict survival time conditioned on the observed covariates. To evaluate the prediction accuracy, we use a generalization of the concordance (c-) index that allows for ties in the event times (Kang *et al.* 2015). The c-index is a non-parametric order statistic that evaluates the pairwise ordering prediction accuracy. A c-index of 1 signifies perfect accuracy, and 0.5 signifies random guessing. Additional details on the c-index are included in Sec. SM-6.

FA-ECPH-C requires model selection for the dimension of the latent space $d_z$. We take a cross-validated approach to model selection. After reserving 25% of each dataset as a test set, we use the remaining 75% of the data for cross-validated model selection with the c-index as the evaluation statistic. The cross validation (CV) and model selection procedure is described in Sec. SM-7 and Sec. SM-8. We search through a small range of dimensions for the latent space, $d_z = \{2, 3, 4, 5\}$ for three real datasets and four simulated datasets and $d_z = \{2, 5, 10, 15\}$ for a fourth real dataset. After choosing the best cross-validated model according c-index, we find the MLE estimates for the parameters trained on the 75% CV set. To evaluate model performance, we compare the c-index from predictions for the 25% test set from FA-ECPH-C with the c-index from predictions from two alternate models. The first alternate model is ECPH-C-$L_1$, with the same observed covariates $\mathbf{x}^{(1)}, \ldots, \mathbf{x}^{(D)}$ used for FA-ECPH-C. The second is a gold standard ECPH-C model, where the gold standard is the clinically accepted best predictors for that disease. Note that both the FA-ECPH-C and ECPH-C-$L_1$ models are at a disadvantage to the gold standard. The gold standard predictors were typically determined with the benefit of additional data unseen by the FA-ECPH-C and ECPH-C-$L_1$ models. Model selection and results for each application are described in the relevant subsections of Sec. 3.

## 2.4 Simulation strategy

To evaluate the performance of the proposed approach, we simulate time-to-event $\tilde{t}$, censoring $\delta$, and covariate data $\mathbf{x}^{(d)}$ from the FA-ECPH-C model. The FA-ECPH-C model is fully generative, and simulating data has a clear procedure. After choosing a model and model parameters, one simulates the latent variable $\mathbf{z}$ from Eqn. (7). Given $\mathbf{z}$, one simulates $\mathbf{x}^{(d)}$ according to Eqn. (5) and (6), and $t$ and $c$ according to Eqn. (1) and (3). The time-to-event and censoring variables are determined by $\tilde{t} = \min(t, c)$ and $\delta = \mathbf{I}(\tilde{t} = t)$. To provide the most stringent test of the model, the test samples are all uncensored; for these samples only $t$ is simulated.

## 3. RESULTS

## 3.1 Application to datasets

We apply our methodology to four datasets with diverse observed covariates. Our results are based upon data collected by the TCGA Research Network (http://cancergenome.nih.gov/), and hosted by the Broad Institutes GDAC Firehose (Broad Institute of MIT and Harvard. 2016); we download the data using the $R$ tool $TCGA2STAT$ developed by Wan *et al.* (2016). $TCGA2STAT$ imports the latest available version-stamped standardized Level 3 dataset on Firehose. The preprocessing consists of a variance filter and a missingness filter for all non-survival datatypes. We accepted only the most variable components across samples up to a percentage that depends on the dataset. Missingness in a covariate was tolerated up in up to 10% of the total samples. We replace the missing values with the sample mean for that covariate. The mRNA and microRNA data is normalized. DNA copy number (variation and alteration) has an additional pre-processing step; the segmentation data reported by TCGA is turned into copy number using the $R$ tool $CNtools$ developed by Zhang (2015) and imbedded in $TCGA2STAT$. The mutation data also has an additional preprocessing step; it is filtered based on status and variant classification and then aggregated at the gene level (Wan *et al.* 2016). Mutation data is modeled as conditionally binomial with $b = 1$; the other non-survival datatypes are modeled as conditionally normal. For some datasets we omit an available datatype because including it would prohibitively reduce the number of samples. If the survival time was reported as zero, we replace the zero value with 1/10 the smallest non-zero survival time. Table 1 contains the different datatypes, the dimension of each type of covariate, and the pre-processing variance filter.

### 3.1.1 Diffuse Lower-Grade Glioma.

Diffuse lower-grade gliomas (LGG) are infiltrative brain tumors that occur most frequently in the cerebral hemisphere of adults. Patients with this diagnosis have highly heterogeneous molecular phenotype, and survival responses. The Cancer Genome Atlas Research Network (2015) has collected clinical, exome sequence, DNA copy number, DNA methylation, messenger RNA expression, and microRNA expression data for many lower-grade gliomas from adults. Two-hundred and seventy-nine gliomas have complete data including survival time. Seventy-two out of 279 patients (26%) are uncensored for survival time. The data collection and data platforms are discussed in detail in The Cancer Genome Atlas Research Network (2015).

A cross-validated model selection search for FA-ECPH-C over latent dimensions $d_z = \{2, 3, 4, 5\}$ identifies dimension $d_z = 5$ as the best FA-ECPH-C model (Fig. 1). A cross-validated search for sparsity parameters $\gamma \in \{1e3, 1e4, 1e5\}$ for ECPH-C-$L_1$ identifies $\gamma = 1e3$ as the best penalty for this model (Fig. 1). For the ECPH-C gold-standard model, we use as covariates age (in years), histological grade ($\in \{2, 3\}$), extent of resection ($\in \{\text{Biopsy Only} = 1, \text{Subtotal Resection} = 2, \text{Gross Total Resection} = 3\}$), and a three-class categorical molecular phenotype corresponding to $IDH - 1p/19q$ status (IDHmut-codel, IDHmut-non-codel, IDHwt) identified by The Cancer Genome Atlas Research Network (2015) as predictive of favorable clinical outcomes. Among all of the models considered during the cross-validated model selection stage, the gold standard ECPH-C has the best performance (Fig. 1). Note that the comparison is biased in favor of the gold-standard model, because the molecular phenotype $IDH - 1p/19q$ was identified using additional data. The best FA-ECPH-C model is competitive with the gold standard.

At latent dimension $d_z = 2$ for FA-ECPH-C, the latent projection of the training and validation data shows three clearly separated clusters (Fig. 2). The Cancer Genome Atlas Research Network (2015) performed a cluster-of-cluster integrative analysis and noted three clusters that correlate with $IDH-1p/19q$ status, but they did not provide a low-dimensional visualization of the clusters. Fig. 3 shows that the clusters found by FA-ECPH-C $d_z = 2$ also strongly correlate with $IDH - 1p/19q$ status and survival time, with the $IDHmut-non-codel$ being most protective, followed by $IDHmut-codel$, and lastly $IDHwt$. The final $d_z = 5$ projection of the data shows that there is additional unexplained variability within the molecular phenotype that is captured by the latent space most predictive of survival time. See Fig. 4 and 5 for the final training and test set latent projections. This highlights the power of the FA-ECPH-C method for integrative sample visualization to identify sample heterogeneity.

Table 1: Covariates used for FA-ECPH-C and ECPH-C-$L_1$ for four TCGA datasets and four simulated datasets (SIM1-4). We accepted only the most variable components up to a percentage that depends on the application. In the table below, ($filter$) refers to the variance filter. Missingness in a covariate is tolerated up in up to 10% of the total samples.

| Disease | No. Tumors (No. uncen.) | No. mRNA (filter) (platform) | No. miRNA (filter) (platform) | No. Mutations (filter) | No. CNV (filter) | No. CNA (filter) | No. DNA Methyl. (filter) (platform) |
|---|---|---|---|---|---|---|---|
| LGG | 279 (72) | 6151 (0.3) (Seq2) | 314 (0.3) (Seq2) | 1455 (0.3) | 6795 (0.3) | 6785 (0.3) | 11816 (0.03) (450K) |
| GBM | 71 (56) | 6151 (0.3) (Seq2) | – | 2431 (0.3) | 6783 (0.3) | 6786 (0.3) | 7485 (0.3) (27K) |
| LUAD | 172 (72) | 6151 (0.3) (Seq2) | 314 (0.3) (Seq2) | 1388 (0.1) | 6786 (0.3) | 6784 (0.3) | – |
| LUSC | 104 (47) | 6151 (0.3) (Seq2) | – | 1370 (0.1) | 6786 (0.3) | 6785 (0.3) | 7475 (0.3) (27K) |
| SIM1, 2 | 279 (171, 172) | 6151 | 314 | 1455 | 6795 | 6785 | 11816 |
| SIM3, 4 | 71, 279 (49, 194) | 6151 | – | 2431 | 6783 | 6786 | 7485 |

The final test set c-index prediction accuracies are summarized in Table 2. For LGG, all of the models performed similarly and well on the final test set.

Table 2: Final model comparisons for FA-ECPH-C for four TCGA datasets and four simulated dataset (SIM1-4). *Note: Recall that Model 7, the gold standard model, is at an advantage, since the predictors were determined with the benefit of additional data.

| Disease | Model No. | Model Type | Test Set c-index | CV c-index mean | CV c-index standard deviation |
|---|---|---|---|---|---|
| LGG | 3 | FA-ECPH-C $d_z = 5$ | 0.81 | 0.72 | 0.08 |
| | 4 | ECPH-C-$L_1$ $\gamma = 1e3$ | 0.81 | 0.81 | 0.03 |
| | 7* | ECPH-C GS | 0.82 | 0.82 | 0.04 |
| GBM | 0 | FA-ECPH-C $d_z = 2$ | 0.46 | 0.66 | 0.08 |
| | 6 | ECPH-C-$L_1$ $\gamma = 1e5$ | 0.29 | 0.44 | 0.05 |
| | 7* | ECPH-C GS | 0.68 | 0.55 | 0.12 |
| LUAD | 1 | FA-ECPH-C $d_z = 3$ | 0.50 | 0.59 | 0.06 |
| | 4 | ECPH-C-$L_1$ $\gamma = 1e3$ | 0.56 | 0.56 | 0.12 |
| | 7* | ECPH-C GS | 0.62 | 0.69 | 0.04 |
| LUSC | 3 | FA-ECPH-C $d_z = 15$ | 0.51 | 0.60 | 0.10 |
| | 4 | ECPH-C-$L_1$ $\gamma = 1e3$ | 0.46 | 0.75 | 0.08 |
| | 7* | ECPH-C GS | 0.45 | 0.53 | 0.08 |
| SIM1 | 3 | FA-ECPH-C $d_z = 5$ | 0.98 | 0.99 | 0.00 |
| | 5 | ECPH-C-$L_1$ $\gamma = 1e5$ | 0.88 | 0.92 | 0.04 |
| SIM2 | 3 | FA-ECPH-C $d_z = 5$ | 0.98 | 0.99 | 0.00 |
| | 6 | ECPH-C-$L_1$ $\gamma = 1e6$ | 0.93 | 0.94 | 0.03 |
| SIM3 | 2 | FA-ECPH-C $d_z = 4$ | 0.91 | 0.85 | 0.14 |
| | 4 | ECPH-C-$L_1$ $\gamma = 5e4$ | 0.75 | 0.70 | 0.08 |
| SIM4 | 1 | FA-ECPH-C $d_z = 3$ | 0.90 | 0.87 | 0.02 |
| | 4 | ECPH-C-$L_1$ $\gamma = 5e4$ | 0.80 | 0.77 | 0.03 |

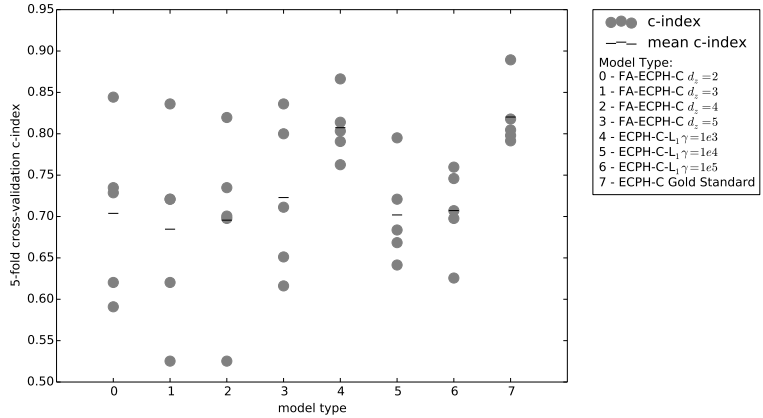

figure 1: Results for the LGG 5-fold CV latent dimension search for FA-ECPH-C, comparison to ECPH-C-$L_1$, and gold standard ECPH-C. Model types are as follows. Models $0-3$ are FA-ECPH-C and have, in order, $d_z = \{2, 3, 4, 5\}$. Models $4-6$ are ECPH-C-$L_1$ and have, in order, $\gamma = \{1e3, 1e4, 1e5\}$, which selects an average of $\{17, 5, 3\}$ relevant covariates. Model 7 is the gold standard ECPH-C model.

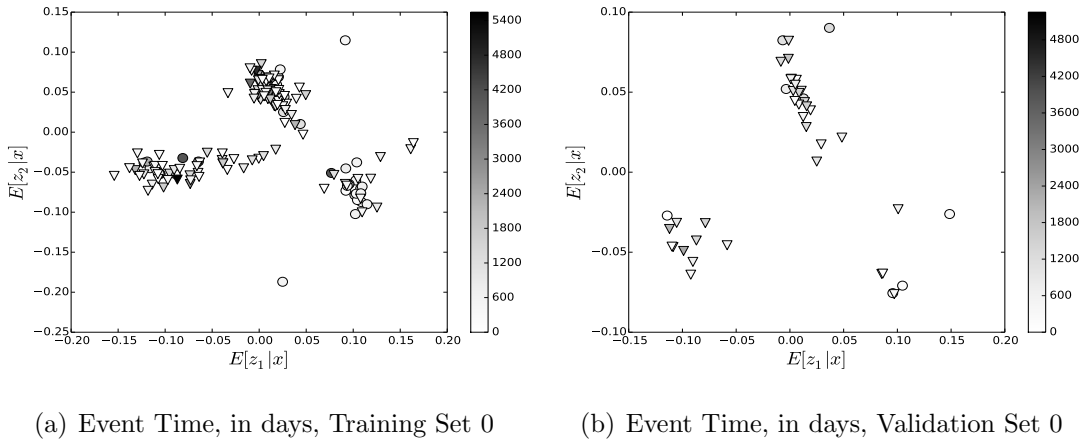

(a) Event Time, in days, Training Set 0

(b) Event Time, in days, Validation Set 0

Figure 2: LGG latent projections $\mathbb{E}[\mathbf{z}|\mathbf{x}^{(1)}, \ldots, \mathbf{x}^{(D)}]$ of the $0^{th}$ CV training and validation cohort for the FA-ECPH-C $d_z = 2$ model. The observations are colored by event time, in days. Circles represent uncensored observations, and triangles represent censored observations.

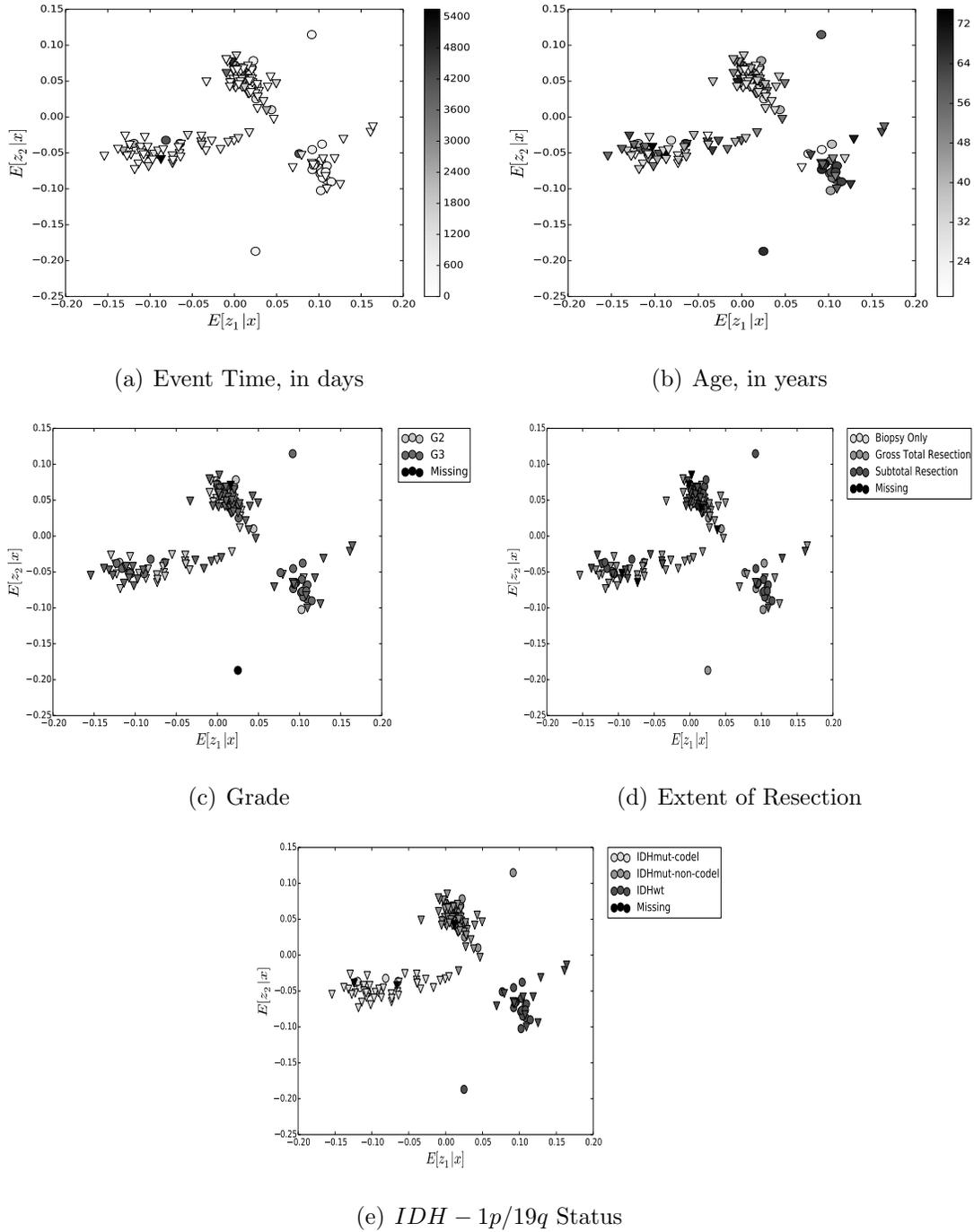

(a) Event Time, in days

(b) Age, in years

(c) Grade

(d) Extent of Resection

(e) $IDH - 1p/19q$ Status

Figure 3: LGG latent projections $\mathbb{E}[\mathbf{z}|\mathbf{x}^{(1)}, \ldots, \mathbf{x}^{(D)}]$ of the $0^{th}$ CV training cohort for the FA-ECPH-C $d_z = 2$ model. Circles represent uncensored observations, and triangles represent censored observations.

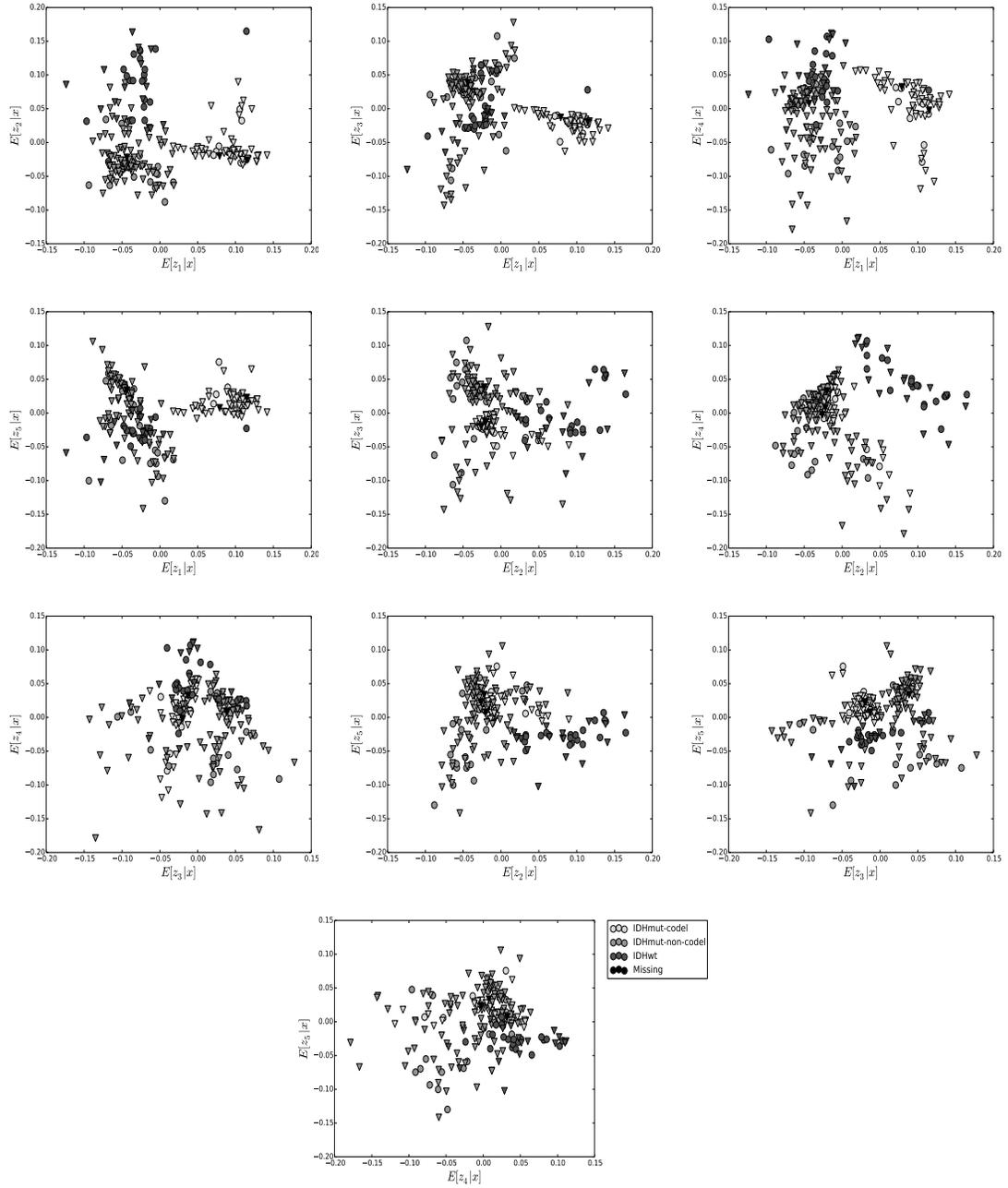

Figure 4: LGG latent projections $\mathbb{E}[\mathbf{z}|\mathbf{x}^{(1)}, \ldots, \mathbf{x}^{(D)}]$ of the training cohort for the FA-ECPH-C $d_z = 5$ model, colored by $IDH-1p/19q$ Status. Circles represent uncensored observations, and triangles represent censored observations.

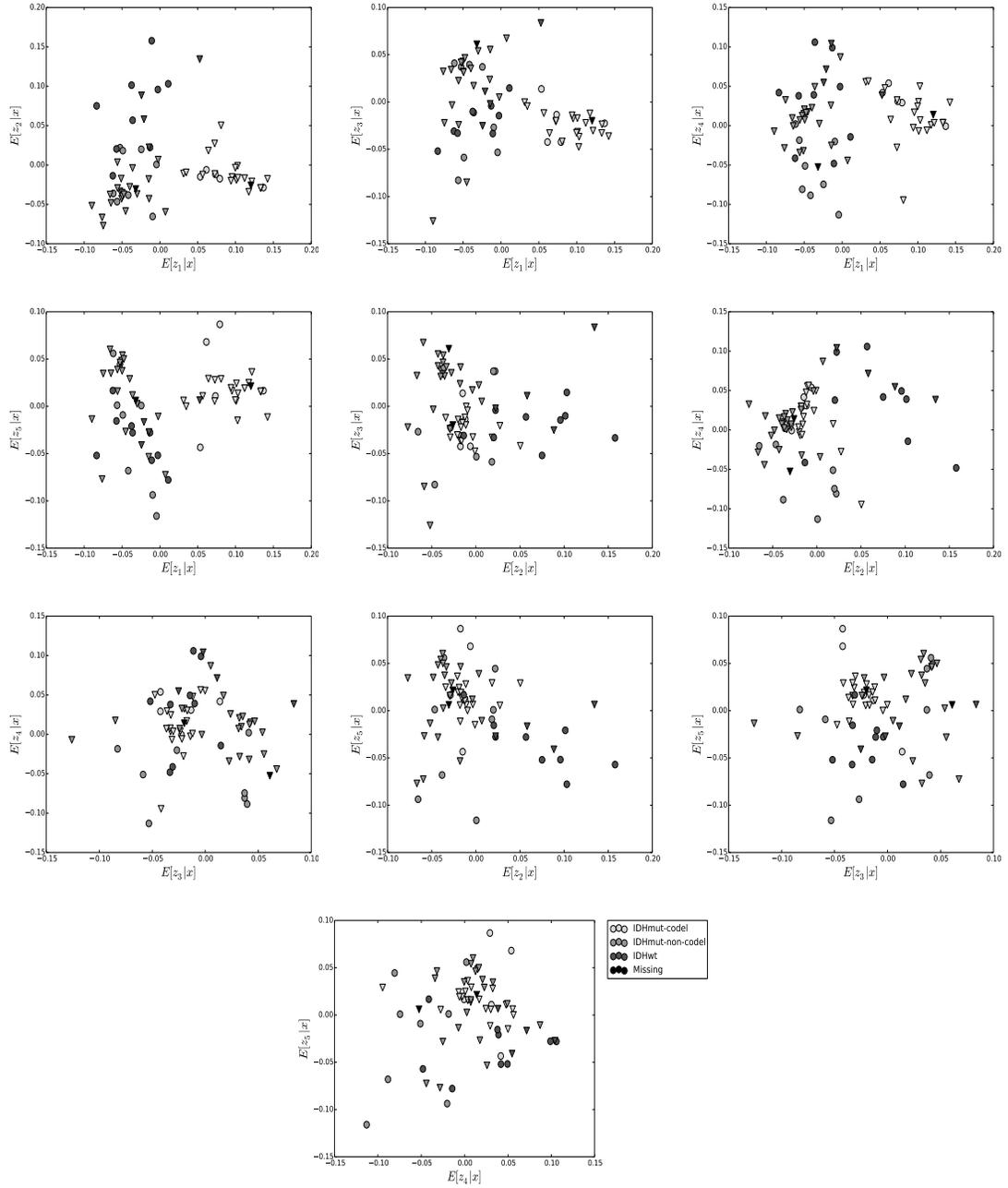

Figure 5: LGG latent projections $\mathbb{E}[\mathbf{z}|\mathbf{x}^{(1)}, \ldots, \mathbf{x}^{(D)}]$ of the test cohort for the FA-ECPH-C $d_z = 5$ model, colored by $IDH - 1p/19q$ Status. Circles represent uncensored observations, and triangles represent censored observations.

### 3.1.2 Glioblastoma multiforme.

Glioblastoma multiforme (GBM) is the most common primary brain tumor in adults. Primary GBM arises *de novo* without progression from previously diagnosed LGG. Like LGG, primary GBM exhibits heterogeneity in molecular phenotype and survival response. Brennan *et al.* (2013) have collected clinical, exome sequence, DNA copy number, DNA methylation, and messenger RNA expression for many glioblastomas from adults. Seventy-one primary glioblastomas have complete data including survival time, and 56 of the 71 patients are uncensored (79%). The data collection and data platforms are discussed in detail in Brennan *et al.* (2013).

A cross-validated search for FA-ECPH-C over latent dimensions $d_z = \{2, 3, 4, 5\}$ identifies latent dimension $d_z = 2$ (Fig. SM-1). The models for $d_z = \{3, 4, 5\}$ are eliminated because for at least one of the five CV sets, the EM algorithm approached a Heywood case (Heywood 1931). A Heywood case has some components of an estimated $\Psi^{(d)}$ approach zero. One of the causes of Heywood cases is too many latent variables. A cross-validated search for sparsity parameters $\gamma \in \{5e3, 1e4, 1e5\}$ for ECPH-C-$L_1$ identifies $\gamma = 1e5$ as the best penalty; however the performance of this model on the GBM data is still quite poor (Fig. SM-1). For the GBM ECPH-C gold-standard model, we include age (in years), a 5-class categorical variable capturing expression subtype (Classical, Mesenchymal, Neural, Proneural, G-CIMP), a binary variable capturing MGMT status ($\in \{methylated = 1, unmethylated = 0\}$), and a binary variable capturing $IDH1$ status ($\in \{WT = 1, R132H = 0\}$) as covariates. These covariates were identified as predictive of positive clinical outcomes in Brennan *et al.* (2013). The performance of the gold standard is worse than the other models considered in the CV model selection stage, despite the fact that the comparison is biased in favor of the gold-standard model. FA-ECPH-C outperforms both the ECPH-C-$L_1$ and gold standard models during CV.

At latent dimension $d_z = 2$ for FA-ECPH-C, the latent projection of the training and validation data shows no appreciable clustering among the patients (Fig. SM-2). Fig. SM-3 shows that the latent projection found by FA-ECPH-C $d_z = 2$ reveals some correlation with the $IDH1$ status, $MGMT$ status, and the expression subtype. The most apparent correlation is with $MGMT$ status.

The final test set c-index prediction accuracy for GBM was quite variable across the final models. The gold standard ECPH-C model performed best, followed by the FA-ECPH-C with $d_z = 2$, and lastly the ECPH-C-$L_1$ with $\gamma = 1e5$. The results are summarized in Table 2.

### 3.1.3 Lung adenocarcinoma

Lung cancer is the leading cause of cancer-related mortality around the world, and lung adenocarcinoma (LUAD) is the most common type of lung cancer. The Cancer Genome Atlas Research Network (2014) collected clinical, exome sequence, DNA copy number, messenger RNA expression, and micro RNA expression data for 172 LUAD tumors, with 72 patients uncensored (42%). See The Cancer Genome Atlas Research Network (2014) for a comprehensive discussion of the data collection and platforms.

A cross-validated search for FA-ECPH-C over latent dimensions $d_z = \{2, 3, 4, 5\}$ identifies dimension $d_z = 3$ (Fig. SM-4). A cross-validated search for sparsity parameters $\gamma \in \{1e3, 1e4, 1e5\}$ for ECPH-C-$L_1$ identifies $\gamma = 1e3$ as the best penalty (Fig. SM-4). For the LUAD gold standard model, we take a 3-class categorical variable for expression subtype (Terminal respiratory unit (TRU), Proximal-proliferative (PP), Proximal-inflammatory (PI)), and a four-class categorical variable for the pathology N-stage (n0, n1, n2, nx). The Cancer Genome Atlas Research Network (2014) found that these covariates were associated with survival outcome, with the TRU subtype exhibiting superior outcomes. This gold standard model is the best predictor of survival time at the CV stage (Fig. SM-4).

The latent projection of the training and validation data shows no apparent clustering among the patients at latent dimension $d_z = 2$ for FA-ECPH-C (Fig. SM-8). The latent projection found by FA-ECPH-C $d_z = 2$, Fig. SM-9, exhibits some correlation with N-stage and expression subtype. The most appreciable correlation is with the classical expression subtype.

The final test set c-index prediction accuracy for LUAD was quite poor across the final models. The gold standard ECPH-C model performed best, followed by the ECPH-C-$L_1$ with $\gamma = 1e3$, and lastly by the FA-ECPH-C with $d_z = 3$. The results are summarized in Table 2.

### 3.1.4 Lung squamous cell carcinoma.

Lung squamous cell carcenoma (LUSC) is the second most common type of lung cancer. The Cancer Genome Atlas Research Network (2012) collected clinical, exome sequence, DNA copy number, DNA methylation, and messenger RNA expression for 104 LUSC tumors. Of the 104 patients, 47 are uncensored (45%). See The Cancer Genome Atlas Research Network (2012) for a comprehensive discussion of the data collection and platforms.

A cross-validated search for FA-ECPH-C over latent dimensions $d_z = \{2, 5, 10, 15\}$ identifies dimension $d_z = 15$ (Fig. SM-7). We search over larger latent dimensions for LUSC than for the other datasets because for LUSC, fitting the larger dimensional models do not result in Heywood cases. A cross-validated search for sparsity parameters $\gamma \in \{1e3, 1e4, 1e5\}$ for ECPH-C-$L_1$ identifies $\gamma = 1e3$ as the best penalty; this is the best performing model for CV (Fig. SM-7). The Cancer Genome Atlas Research Network (2012) does not perform a survival analysis on the cohort, and they do not identify specific covariates as predictive of positive clinical outcomes. We follow Jiang *et al.* (2014) and use gender, smoking history, age at initial pathologic diagnosis, and tumor stage as covariates in our gold standard model. In particular, we use a binary variable for gender ($\in \{Female = 1, Male = 0\}$), a 4-class categorical variable for smoking history (Lifelong non-smoker, Current reformed smoker for $\leq 15$ years, Current reformed smoker for $> 15$ years, and Current smoker), and a 7-class categorical variable for pathology T-stage (t1, t1a, t1b, t2, t2a, t3, and t4). This gold standard model is a poor predictor of survival time (Fig. SM-7). FA-ECPH-C outperforms the gold standard model during CV.

The latent projection of the training and validation data shows no apparent clustering among the patients at latent dimension $d_z = 2$ for FA-ECPH-C (Fig. SM-8). The latent projection found by FA-ECPH-C $d_z = 2$, Fig. SM-9, exhibits some correlation with the smoking status, gender, and T-stage. The most appreciable correlation is with T-stage.

The final test set c-index prediction accuracy for LUSC was quite poor across the final models. The FA-ECPH-C with $d_z = 15$ model performed best, followed by the ECPH-C-$L_1$ with $\gamma = 1e3$, and last by the gold standard ECPH-C model. The results are summarized in Table 2.

## 3.2 Simulation study

We test the efficacy of our model via simulation through four studies, SIM1-4. Table 2 contains the c-index performance for all four simulation studies. We search across $d_z = \{2, 3, 4, 5\}$ and $\gamma \in \{5e4, 1e5, 1e6\}$ during CV for SIM1-4. For SIM1, we take the final parameters learned for the $d_z = 5$ FA-ECPH-C model on the LGG dataset and simulate data for 279 samples according to the model. We select simulation parameters in this manner so that our simulation study is as close to real data as possible. Five-fold CV for FA-ECPH-C on the training samples selects the Model 3 ($d_z = 5$) as the best model. Cross validation for ECPH-C-$L_1$ for the sparsity parameter identifies $\gamma = 1e5$ as the best sparsity penalty for this model. In this simulation, FA-ECPH-C performs very well, and outperforms ECPH-C-$L_1$ in CV and on the final test set. In SIM2, we repeat the simulation and analysis to assess the stability. We find that the FA-ECPH-C CV and test set results are stable, and the ECPH-C-$L_1$ results are less stable.

For our last two simulation studies, SIM3 and SIM4, we take the final parameters learned for the $d_z = 2$ FA-ECPH-C model on the GBM dataset. For SIM3, we simulate data for 71 samples according to the model. We perform this simulation to show the prediction accuracy on the smallest of sample sizes and with a different distribution of parameters. Model 2 ($d_z = 4$) is selected by five-fold CV for FA-ECPH-C, although all models perform similarly. The sparsity parameter $\gamma = 5e4$ is selected for ECPH-C-$L_1$. FA-ECPH-C outperforms ECPH-C-$L_1$ during CV and on the final test set for SIM3. The c-index performance for SIM3 is not as good as SIM1-2, and the SIM 3 CVs standard deviations are larger. We perform SIM4 to assess whether this difference is due to the sample size or the parameters. In SIM4, we simulate 279 samples with the same parameters as SIM3. Model 1 ($d_z = 3$) is selected by five-fold CV for FA-ECPH-C, and the sparsity parameter $\gamma = 5e4$ is selected for ECPH-C-$L_1$. The CV c-indices for model selection for SIM4 are shown in Fig. 6. We find that the c-index performance for FA-ECPH-C SIM4 is equivalent to SIM3, but with improved CV standard deviation. The performance and standard deviation for ECPH-C-$L_1$ both improve with the larger sample size. The contrast between SIM1-2 and SIM4 shows that the prediction accuracy of the FA-ECPH-C model depends on the model parameters.

We illustrate the latent projection with SIM4. We assign the 279 samples to three groups according to their true latent coordinates (Fig. 7). See Fig. 8 and 9 for latent projections of the learning and test sets for the final model for SIM4. The learned and predicted latent spaces keep the three groupings of points proximate.

These simulation studies shows that, on data generated by the FA-ECPH-C model, the FA-ECPH-C model can both successfully predict survival time orderings among samples and outperform the alternative model ECPH-C-$L_1$ on datasets with these small sample sizes, large covariate dimensions, and heavy censoring characteristics. We also show that the model performance depends on the model parameters.

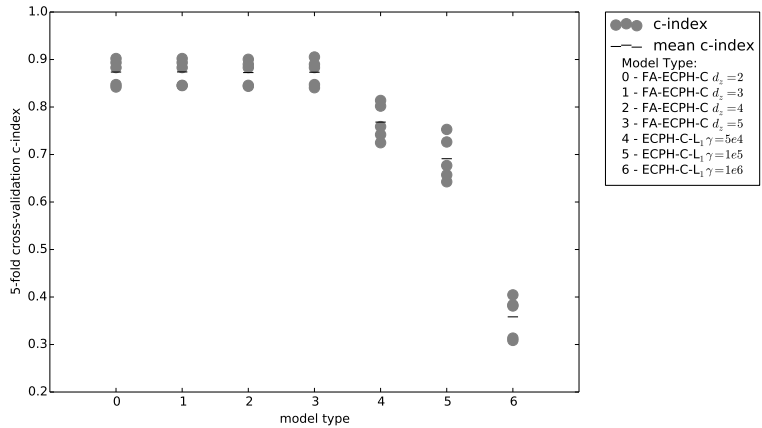

Figure 6: Results for the SIM4 5-fold CV latent dimension search for FA-ECPH-C and comparison to the ECPH-C-$L_1$ model. Model types are as follows. Models $0-3$ are FA-ECPH-C and have, in order, $d_z = \{2, 3, 4, 5\}$. Models $4-6$ are ECPH-C-$L_1$ and have, in order, $\gamma = \{5e4, 1e5, 1e6\}$, which selects an average of $\{5, 5, 5\}$ relevant covariates.

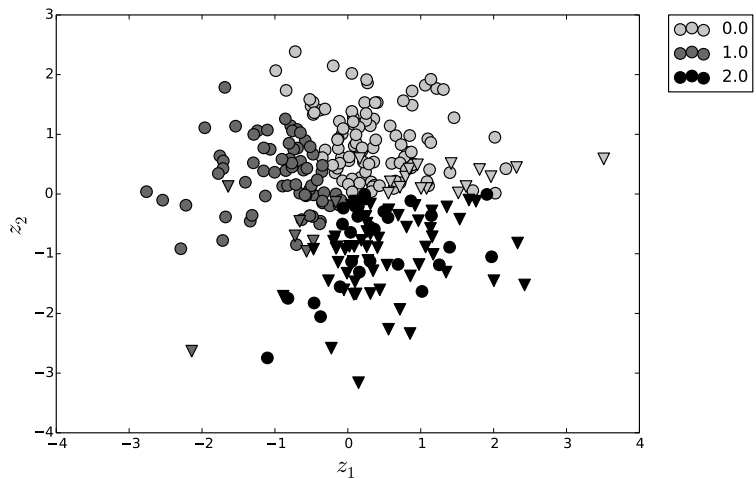

Figure 7: All of the SIM4 samples in the true latent space **z**.

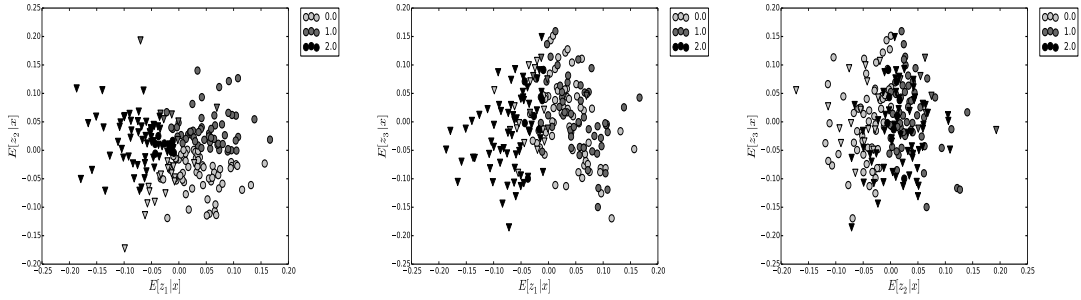

Figure 8: SIM4 latent projections $\mathbb{E}[\mathbf{z}|\mathbf{x}^{(1)}, \ldots, \mathbf{x}^{(D)}]$ of the training cohort for the FA-ECPH-C $d_z = 3$ model, colored by group membership. Circles represent uncensored observations, and triangles represent censored observations.

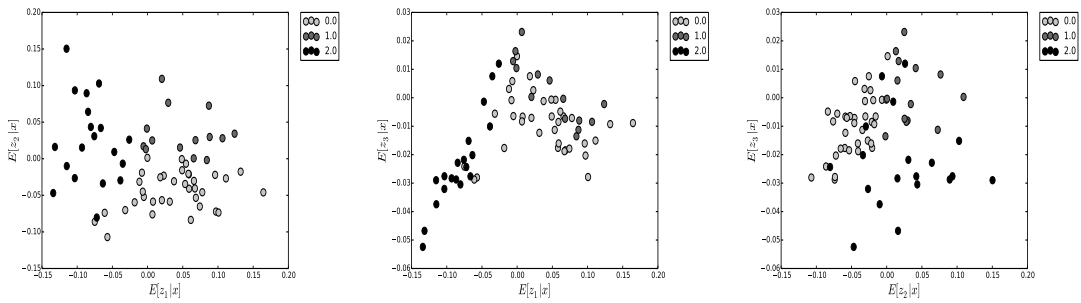

Figure 9: SIM4 latent projections $\mathbb{E}[\mathbf{z}|\mathbf{x}^{(1)}, \ldots, \mathbf{x}^{(D)}]$ of the test cohort for the FA-ECPH-C $d_z = 3$ model, colored by group membership. Circles represent uncensored observations, and triangles represent censored observations.

# 4. DISCUSSION

In this paper, we propose a joint factor model and an exponential Cox proportional hazards model with diverse covariates for survival time prediction for informative censoring in the high-dimensional regime. The latent space of factor analysis are the covariates in the ECPH model for the event of interest and the censoring event. Related models for survival analysis have been previously considered for non-informative censoring (Larsen 2004; Muthn and Masyn 2005; Larsen 2005; Asparouhov and Masyn 2006, Wong *et al.* 2017), in the low-dimensional regime. The prior body of work on latent variable survival models has not considered the high-dimensional setting.

We perform simulation studies to show the efficacy of our method. We show that on data generated by the FA-ECPH-C model with parameters learned from real data, and a sample size commensurate with real data, the FA-ECPH-C model successfully predicts survival time orderings among samples and can outperform the alternative model (ECPH-C-$L_1$). The performance of the method and the alternative depends on the choice of parameters for the model.

This proposed model is competitive for survival time prediction with the two alternative approaches: an $L_1$-penalized exponential Cox proportional hazards model and an unpenalized gold-standard exponential Cox proportional hazards model. On the LUSC dataset, the proposed model outperforms the two alternatives. On the GBM and LUAD datasets the gold standard model outperforms. However, for the GBM, LUAD, and LUSC datasets, the model generally performs poorly, comparable to random guessing. On the LGG dataset, all three models perform approximately equally, and very well. The LGG dataset has the advantage, compared to the other datasets of a much larger sample size, although it is most heavily censored. It also has striking low-dimensional structure correlated with survival time, consistent with the foundational hypothesis of the proposed model. The LGG dataset also illustrates the value of the low-dimensional projection of the data for identifying heterogeneity in the individuals. There are three distinct clusters of patients in the low-dimensional latent space. These clusters correlate with the $IDH-1p/19q$ molecular phenotype identified by The Cancer Genome Atlas Research Network (2015). This is strong evidence that the FA-ECPH-C model can provide a meaningful representation of real data.

The FA-ECPH-C model shows promise. Future work includes applying this model to additional diseases, and revisiting LGG, GBM, LUAD, and LUSC as more data is collected. Future directions for latent variable survival modeling include extending the model to handle missing data, the inclusion of non-informative covariates, and alternate conditional distributions for survival time.

## SUPPLEMENTARY MATERIAL

**Supplementary Materials:** contains technical appendices, additional simulations, and figures. (pdf)

**Github:** Python-package *survival_factor_model* containing code to perform the methods described in the article can be found at https://github.com/srmcc/survival_factor_model.git. The package also contains code to download or generate all datasets used as examples in the article.

# A latent variable model for survival time prediction with censoring and diverse covariates

## Supplementary Materials

Shannon R. McCurdy[2,1]   Annette Molinaro[3]   Lior Pachter[4]

[2]  California Institute for Quantitative Biosciences,
    University of California, Berkeley,
    Berkeley, CA.

[3]  Departments of Neurological Surgery, Epidemiology, and Biostatistics,
    University of California, San Francisco,
    San Francisco, CA.

[4]  Division of Biology and Biological Engineering and
    Department of Computing and Mathematical Sciences,
    California Institute of Technology,
    Pasadena, CA.

# Contents

[1]  Corresponding author. E-mail: smccurdy@berkeley.edu

# 1 A note about notation

Lowercase boldface letters such as $\mathbf{x}$ are vectors. Uppercase boldface letters such as $\mathbf{X}$ are matrices. We reserve the index $n$ as a sample index. The total number of individuals (samples) are $N$. $N$ observations of $\mathbf{x}$ are $\mathbf{X}$. If an uppercase boldface letter has an index, such as $\mathbf{X}_i$, it denotes the $i^{th}$ row or column vector from the matrix $\mathbf{X}$, and the dimension of the vector will be specified in the text.

# 2 The exponential Cox proportional hazards model with censoring and $L_1$ penalty (ECPH-C-$L_1$)

Let $t$, a non-negative random variable, be the survival time of an individual. We denote the non-random $(p \times 1)$-dimensional vector of covariates $\mathbf{x}$ for that individual. If a subset $s' = s + 1$ of the $p' = p + 1$ covariates have a fixed sum (e.g. categorical covariates), we only include $s$ elements of the subset in the $p$ covariates. Under the ECPH model, there are two parameters of interest, the time-independent baseline hazard $\lambda$ and the $(p \times 1)$-dimensional vector of regression coefficients $\boldsymbol{\beta}$. For notational efficiency, we set,

$$\tilde{\mathbf{x}} = \begin{pmatrix} 1 \\ \mathbf{x} \end{pmatrix} \qquad \mathbf{w} = \begin{pmatrix} \ln \lambda \\ \boldsymbol{\beta} \end{pmatrix} \tag{1}$$

Under this model, the survival time distribution, given the covariates $\tilde{\mathbf{x}}$, is the exponential distribution with the following parameterization of the standard exponential rate parameter $\rho$,

$$t|\mathbf{x} \sim \mathrm{Exp}(\rho = \exp\left(\mathbf{w}^T \tilde{\mathbf{x}}\right)) \tag{2}$$

We take the following model for non-informative censoring. There are two non-negative random variables, $t$ and $c$, that represent the survival time and the censoring time of the individual. We take the probability densities for $t$ and $c$ to be given by Eqn. (2), with parameters $\mathbf{w}_T$ and $\mathbf{w}_C$, respectively, and $t \perp c | \mathbf{x}$. In the censored setting, one is not able to observe both $t$ and $c$ for an individual. Instead, the observed random variables for each individual are $\tilde{t} = \min(t, c)$ and $\delta = \mathbf{I}(\tilde{t} = t)$, where $\mathbf{I}$ is the indicator function. The probability of the observed data for an individual is simply the cumulative probability for each event,

$$p_{\mathbf{w}_T, \mathbf{w}_C}(\tilde{t}, \delta | \mathbf{x}) = \left(p_{\mathbf{w}_T}(\tilde{t}|\mathbf{x}) P_{\mathbf{w}_C}(c > \tilde{t}|\mathbf{x})\right)^\delta \left(P_{\mathbf{w}_T}(t > \tilde{t}|\mathbf{x}) p_{\mathbf{w}_C}(\tilde{t}|\mathbf{x})\right)^{1-\delta}. \tag{3}$$

$P_{\mathbf{w}_{T,C}}((t,c) > \tilde{t}|\mathbf{x})$ is the survival function for $(t, c)$, respectively. As a shorthand, we use the subscript $\{T, C\}$ for generic expressions for the respective $(t, c)$ distributions.

We assume that the individual observations of the data are independent, so the likelihood of the data of $N$ individuals is simply the product of the individual probabilities. The likelihood of the data can be rearranged into a product of a partial likelihoods for $\mathbf{w}_T$ and $\mathbf{w}_C$,

$$\begin{aligned} \mathcal{L}(\mathbf{w}_T, \mathbf{w}_C) &= \mathcal{L}(\mathbf{w}_T) \mathcal{L}(\mathbf{w}_C) \\ \mathcal{L}(\mathbf{w}_{T,C}) &= \prod_{n=1}^{N} p_{\mathbf{w}_{T,C}}(\tilde{t}_n | \mathbf{x}_n)^{\delta_n^{T,C}} P_{\mathbf{w}_{T,C}}((t,c) > \tilde{t}_n | \mathbf{x}_n)^{1-\delta_n^{T,C}} \end{aligned} \tag{4}$$

where we have introduced $\delta^T = \delta$ and $\delta^C = 1 - \delta$.

If an $L_1$-penalty is desired, or for the setting of $p > N$, required, the partial log-likelihoods become,

$$\ln\left(\mathcal{L}_{L_1}(\mathbf{w}_{T,C})\right) = \ln\left(\mathcal{L}(\mathbf{w}_{T,C})\right) - \gamma_{T,C} |\mathbf{w}_{T,C}|, \tag{5}$$

where $\gamma_{T,C}$ sets the degree of sparsity induced by the penalty. Determining the appropriate $\gamma_{T,C}$ is a model selection problem.

## 2.1 Inference

The parameters in the ECPH model can be estimated through an iterative maximum likelihood procedure. Following the insight by Tibshirani (1997), the maximum likelihood estimates,

$$\hat{\mathbf{w}}_{T,C} = \max_{\mathbf{w}_{T,C}} \mathcal{L}(\mathbf{w}_{T,C})$$

can be re-framed as an iterative least-squares minimization problem through a first-order Taylor expansion of $\mathcal{L}(\mathbf{w}_{T,C})$. For notational convenience, arrange the data for each individual $\tilde{\mathbf{x}}_n$ into a $(p+1) \times N$-dimensional matrix $\tilde{\mathbf{X}}$, and $\delta_n, \tilde{t}_n$ into $1 \times N$-dimensional vectors $\boldsymbol{\delta}, \tilde{\mathbf{t}}$. Another convenient definition is $\boldsymbol{\eta}_{T,C} = \mathbf{w}_{T,C}^T \tilde{\mathbf{X}}$. The minimization problem for $\hat{\mathbf{w}}_{T,C}^{(s+1)}$ at step $s+1$ is,

$$\hat{\mathbf{w}}_{T,C}^{(s+1)} = \min_{\mathbf{w}_{T,C}^{(s+1)}} \left\| \left( \boldsymbol{\eta}_{T,C}^{(s)} + \boldsymbol{\delta}^{T,C} \circ \tilde{\mathbf{t}}^{-1} \circ \exp\left(-\boldsymbol{\eta}_{T,C}^{(s)}\right) - \mathbb{1} - \left(\mathbf{w}_{T,C}^{(s+1)}\right)^T \tilde{\mathbf{X}} \right) \mathrm{diag}\left(\sqrt{\tilde{\mathbf{t}}} \circ \exp(\tfrac{1}{2}\boldsymbol{\eta}_{T,C}^{(s)})\right) \right\|^2$$

where $\circ$ is the element-wise Hadamard product, $(\exp(\mathbf{x}), \mathbf{x}^{-1})$ are applied element-wise, and $\mathbb{1}$ is a $1 \times N$ vector of ones. We initialize as follows,

$$\mathbf{w}_{T,C}^{(0)} = \begin{pmatrix} \ln\left(\frac{\sum_{n=1}^N \delta_n^{T,C}}{\sum_{n=1}^N \tilde{t}_n}\right) \\ \mathbb{0} \end{pmatrix},$$

where $\mathbb{0}$ is a $p \times 1$ vector of zeros.

If an $L_1$ penalized solution for $\hat{\mathbf{w}}_{T,C}$ is desired, use standard algorithms such as least-angle-regression-lasso (LARS-lasso) (Efron *et al.* 2004) at each iteration (Tibshirani 1997).

We find $\hat{\mathbf{w}}_{T,C}$ converges quickly; we use 5 iterations.

## 2.2 Prediction

Once we have the maximum likelihood estimate for the regression coefficients, $\hat{\mathbf{w}}_T$, we can calculate the expectation of the time-to-event given the covariates $\mathbf{x}$ for an individual. Under the ECPH model (Eqn. 2), the prediction $\hat{t}$ for the individual's time-to-event is,

$$\hat{t} = \mathbb{E}_{\hat{\mathbf{w}}_T}[t|\mathbf{x}] = \exp(-\hat{\mathbf{w}}_T^T \tilde{\mathbf{x}}).$$

## 3 Factor analysis (FA)

For a comprehensive review of FA, see Bartholomew *et al.* (2011) and references therein. In the simplest form, FA is the following linear, Gaussian latent variable model. For $\mathbf{z}$ a $(d_z \times 1)$-dimensional latent random vector, $\mathbf{x}$ a $(d_x \times 1)$-dimensional observed random vector of covariates (with $d_x \geq d_z$), $\mathbf{W}$ a $(d_x \times d_z)$ matrix of factor loading parameters, $\boldsymbol{\mu}$ a $(d_x \times 1)$-dimensional vector of mean parameters, and $\boldsymbol{\Psi} = \mathrm{diag}(\boldsymbol{\psi})$ a $(d_x \times d_x)$ diagonal matrix of error parameters,

$$\begin{aligned} \mathbf{z} &\sim \mathcal{N}(\mathbb{0}, \mathbb{1}) \\ \mathbf{x}|\mathbf{z} &\sim \mathcal{N}(\mathbf{W}\mathbf{z} + \boldsymbol{\mu}, \boldsymbol{\Psi}) \end{aligned} \qquad (6)$$

Here, $\mathbb{0}$ is a $d_z$ or $d_x$ -dimensional zero vector and $\mathbb{1}$ is the $d_z$-dimensional identity matrix. Note that because $\mathbf{\Psi}$ is diagonal, the individual components of $\mathbf{x}$ are conditionally independent of each other, given $\mathbf{z}$. The joint probability density of the observed covariates and the latent space is simply,

$$p(\mathbf{x}, \mathbf{z}) = p(\mathbf{x}|\mathbf{z})p(\mathbf{z}). \tag{7}$$

From the properties of multivariate Gaussians, the marginal distribution of the observed covariates is,

$$\mathbf{x} \quad \sim \quad \mathcal{N}(\boldsymbol{\mu}, \mathbf{W}\mathbf{W}^T + \mathbf{\Psi}). \tag{8}$$

From (Eqn. 8), we see that the unobserved latent space is responsible for the covariance of the observed covariates $\mathbf{x}$ through the lower-rank matrix product of factor loadings $\mathbf{W}\mathbf{W}^T$ and the individual variances $\mathbf{\Psi}$. This lower- rank approximation to the covariance matrix of $\mathbf{x}$ reduces the total number of degrees of freedom needed to estimate the covariance and is a form of regularization. The factor loadings and the latent variables are only identifiable up to an orthogonal transformation. Determining the unknown dimension $d_z$ of the latent space is a model selection problem.

We assume that the individual observations of the data are independent, so the complete likelihood of $N$ individuals is simply the product of the individual probabilities (Eqn. 7).

## 3.1 Additional conditional distributions

One can generalize the distributions considered for $\mathbf{x}|\mathbf{z}$ using the framework of generalized linear models, such as Tipping (1999) for the Bernoulli distribution, and more distributions in Wedel and Kamakura (2001). See McCullagh and Nelder (1989) for a discussion of generalized linear models. In this framework, the natural parameter of the conditional distribution of $\mathbf{x}|\mathbf{z}$ is taken to be linear in the latent space $\mathbf{W}\mathbf{z} + \boldsymbol{\mu}$. For the purpose of this note, we will consider a high-dimensional collection of conditionally binomial covariates and also conditionally multinomial covariates. For the $d_x$ conditionally binomial covariates, we have,

$$\mathbf{x}|\mathbf{z} \sim \text{Binomial}(b, \mathbf{f} = \sigma(\mathbf{W}\mathbf{z} + \boldsymbol{\mu})). \tag{9}$$

Here, $\sigma(x) = (1 + \exp(-x))^{-1}$ is the logistic function, and it is applied element-wise. The Bernoulli distribution has $b = 1$.

For a conditionally multinomial covariate of dimension $d_x$, we have,

$$\mathbf{x}|\mathbf{z} \sim \text{Multinomial}(b, \mathbf{f} = \text{softmax}(\mathbf{W}\mathbf{z} + \boldsymbol{\mu})). \tag{10}$$

Here, the softmax function is a generalization of the logistic function to a $d_x \times 1$-dimensional vector-valued function,

$$\text{softmax}(\mathbf{x}) = \frac{\exp(\mathbf{x})}{\sum_{i=1}^{d_x} \exp(x_i)}, \tag{11}$$

where $\exp(\mathbf{x})$ is applied element-wise to the vector $\mathbf{x}$. To have the appropriate number of parameters for the multinomial distribution, we set the elements of the last row of $\mathbf{W}$ and $\boldsymbol{\mu}$ equal to zero. The categorical distribution has $b = 1$. When $d_x = 2$, we recover a single conditionally binomial distribution.

## 3.2 Diverse conditional distributions

One can consider FA with diverse conditional distributions on the observed covariates. By this we mean that a subset of the observed covariates $\mathbf{x}^{(1)}$ may be conditionally normal, while a different subset $\mathbf{x}^{(d)}$ may be conditionally binomial, and so on for the $D$ different datatypes. For example,

$$\begin{aligned}
\mathbf{x}^{(1)}|\mathbf{z} &\sim \mathcal{N}(\mathbf{W}^{(1)}\mathbf{z} + \boldsymbol{\mu}^{(1)}, \boldsymbol{\Psi}^{(1)}) \\
&\ldots \\
\mathbf{x}^{(d)}|\mathbf{z} &\sim \text{Binomial}(b^{(d)}, \mathbf{f}^{(d)} = \sigma(\mathbf{W}^{(d)}\mathbf{z} + \boldsymbol{\mu}^{(d)})). \\
&\ldots \\
\mathbf{x}^{(D)}|\mathbf{z} &\sim \text{Multinomial}(b^{(D)}, \mathbf{f}^{(D)} = \text{softmax}(\mathbf{W}^{(D)}\mathbf{z} + \boldsymbol{\mu}^{(D)})).
\end{aligned} \tag{12}$$

In this setting, the joint probability of the observed covariates and the latent variables is,

$$p_\Theta(\mathbf{x}^{(1)}, \ldots, \mathbf{x}^{(D)}, \mathbf{z}) = \left(\prod_{d=1}^{D} p_{\Theta^{(d)}}(\mathbf{x}^{(d)}|\mathbf{z})\right) p(\mathbf{z}), \tag{13}$$

where we have collected all of the parameters for the $d^{th}$ conditional distribution into $\Theta^{(d)}$ and all of the parameters of the joint distribution into $\Theta = \{\Theta^{(1)}, \ldots, \Theta^{(D)}\}$.

## 3.3 Inference

There is no closed-form solution for the maximum likelihood estimates of the parameters of FA, but the estimates can be obtained through the Expectation-Maximization (EM) algorithm (Dempster *et al.* 1977) applied to the complete log-likelihood (Rubin and Thayer 1982). For notational convenience, arrange the data for each individual $\mathbf{x}_n$ into a $d_x \times N$-dimensional matrix $\mathbf{X}$, $\mathbf{z}_n$ into a $d_z \times N$-dimensional matrix $\mathbf{Z}$, and $\boldsymbol{\mu}$ is a $d_x \times N$ matrix with each column identical. We follow Rubin and Thayer (1982) and find $\hat{\boldsymbol{\mu}}$ from the maximum likelihood of the marginal likelihood of the observed covariates,

$$\hat{\boldsymbol{\mu}} = \frac{1}{N}\sum_{n=1}^{N}\mathbf{x}_n. \tag{14}$$

From the complete log-likelihood, then, the expectation (E) -step, given parameters $\boldsymbol{\mu}, \mathbf{W}, \boldsymbol{\Psi}$ is,

$$\mathbb{E}[\mathbf{Z}|\mathbf{X}] = (\mathbf{W}^T\boldsymbol{\Psi}^{-1}\mathbf{W} + \mathbb{1})^{-1}\mathbf{W}^T\boldsymbol{\Psi}^{-1}(\mathbf{X} - \boldsymbol{\mu}) \tag{15}$$

$$\mathbb{E}[\mathbf{Z}\mathbf{Z}^T|\mathbf{X}] = N(\mathbf{W}^T\boldsymbol{\Psi}^{-1}\mathbf{W} + \mathbb{1})^{-1} + \mathbb{E}[\mathbf{Z}|\mathbf{X}]\mathbb{E}[\mathbf{Z}|\mathbf{X}]^T. \tag{16}$$

The maximization (M) -step of the parameters is,

$$\hat{\mathbf{W}}|\hat{\boldsymbol{\mu}} = (\mathbf{X} - \hat{\boldsymbol{\mu}})\,\mathbb{E}[\mathbf{Z}|\mathbf{X}]^T\mathbb{E}[\mathbf{Z}\mathbf{Z}^T|\mathbf{X}]^{-1}$$
$$\hat{\boldsymbol{\Psi}}|\hat{\boldsymbol{\mu}}, \hat{\mathbf{W}} = \frac{1}{N}\mathrm{diag}\left((\mathbf{X} - \hat{\boldsymbol{\mu}})(\mathbf{X} - \hat{\boldsymbol{\mu}})^T - \hat{\mathbf{W}}\mathbb{E}[\mathbf{Z}\mathbf{Z}^T|\mathbf{X}]\hat{\mathbf{W}}^T\right). \tag{17}$$

### 3.3.1 Additional conditional distributions

The conditional binomial distribution, Binomial$(b, \mathbf{f} = \sigma(\mathbf{W}\mathbf{z} + \boldsymbol{\mu}))$, is,

$$p_{b,\mathbf{W},\boldsymbol{\mu}}(\mathbf{x}|\mathbf{z}) = \left(\prod_{i=1}^{d_x}\binom{b}{x_i}\sigma^b(-(\mathbf{W}_i\mathbf{z} + \mu_i))\right)\exp\left(\mathbf{x}^T(\mathbf{W}\mathbf{z} + \boldsymbol{\mu})\right) \tag{18}$$

where the elements of $\mathbf{x}$ are $x_i \in \{0, 1, \ldots, b\}$. This conditional distribution has a drawback in that the marginal distribution of the observed covariates, $p(\mathbf{x})$, is intractable. Using the insight of Jaakkola and Jordan (1997) and Tipping (1999), we introduce a variational approximation to the logistic function $\sigma$ in (Eqn. 18). This approximation is,

$$\sigma(x) \geq \sigma(\xi)\exp\left(\frac{1}{2}(x - \xi) - \lambda(\xi)(x^2 - \xi^2)\right) \qquad \lambda(\xi) = \frac{1}{2\xi}\left(\sigma(\xi) - \frac{1}{2}\right), \tag{19}$$

where $\xi$ is the variational parameter. Note that the approximation is exact when $\xi = x$. Under this approximation, $p_{b,\mathbf{W},\boldsymbol{\mu}}(\mathbf{x}|\mathbf{z})$ is approximated by $\tilde{p}_{b,\mathbf{W},\boldsymbol{\mu},\boldsymbol{\xi}}(\mathbf{x}|\mathbf{z})$, where there is a $d_x \times 1$ vector of variational parameters $\boldsymbol{\xi}$ (for $N$ individuals there are $d_x \times N$),

$$\tilde{p}_{b,\mathbf{W},\boldsymbol{\mu},\boldsymbol{\xi}}(\mathbf{x}|\mathbf{z}) = \left(\prod_{i=1}^{d_x}\binom{b}{x_i}\sigma^b(\xi_i)\exp\left(-\frac{b}{2}(\mathbf{W}_i\mathbf{z} + \mu_i + \xi_i) - b\lambda(\xi_i)((\mathbf{W}_i\mathbf{z} + \mu_i)^2 - \xi_i^2)\right)\right)$$
$$\exp\left(\mathbf{x}^T(\mathbf{W}\mathbf{z} + \boldsymbol{\mu})\right). \tag{20}$$

This approximation is a lower bound: $\tilde{p}_{b,\mathbf{W},\boldsymbol{\mu},\boldsymbol{\xi}}(\mathbf{x}|\mathbf{z}) \leq p_{b,\mathbf{W},\boldsymbol{\mu}}(\mathbf{x}|\mathbf{z})$ and $\tilde{p}_{b,\mathbf{W},\boldsymbol{\mu},\boldsymbol{\xi}}(\mathbf{x}) \leq p_{b,\mathbf{W},\boldsymbol{\mu}}(\mathbf{x})$. An additional benefit is that the variational approximation is quadratic in $\mathbf{z}$. As a consequence, the EM updates for the approximate complete log-likelihood are analytic, and the conditional probability of $\tilde{p}(\mathbf{z}|\mathbf{x})$ becomes Gaussian.

With this approximation, for individuals $n \in \{1, \ldots, N\}$, the E-steps are,

$$\mathbb{E}[z_{jn}|\mathbf{x}_n] = \sum_{k=1}^{d_z}(C_n)_{jk}\left(\sum_{i=1}^{d_x} W_{ik}\left(x_{in} - \frac{b}{2} - 2b\lambda(\xi_{in})\mu_{in}\right)\right) \quad (21)$$

$$\mathbb{E}[z_{jn}z_{kn}|\mathbf{x}_n] = (C_n)_{jk} + \mathbb{E}[z_{jn}|\mathbf{x}_n]\mathbb{E}[z_{kn}|\mathbf{x}_n]$$

$$\text{where } ((C_n)^{-1})_{jk} = \left(\delta_{jk} + 2b\sum_{i=1}^{d_x}\lambda(\xi_{in})W_{ij}W_{ik}\right) \quad (22)$$

In these expressions, repeated indices do not imply summations. $\mu_{in}$ is the $(in)^{th}$ element of the matrix of $N$ columns of the $d_x \times 1$ vector $\boldsymbol{\mu}$. $\delta_{jk}$ is the Kronecker delta. Each $\mathbf{C}_n$ is a $d_z \times d_z$ matrix.

The M-steps are,

$$\hat{\xi}_{in}^2|W_{ij},\mu_{in} = \sum_{k=1}^{d_z}\sum_{j=1}^{d_z}W_{ij}W_{ik}\mathbb{E}[z_{jn}z_{kn}|\mathbf{x}_n] + 2\sum_{j=1}^{d_z}W_{ij}\mathbb{E}[z_{jn}|\mathbf{x}_n]\mu_{in} + \mu_{in}^2$$

$$\hat{W}_{ik}|\hat{\xi}_{in},\mu_{in} = \min_{W_{ik}}\frac{1}{2}\left(\sum_{j=1}^{d_z}\sum_{n=1}^{N}\left(x_{in} - \frac{b}{2} - 2b\lambda(\hat{\xi}_{in})\mu_{in}\right)\mathbb{E}[z_{jn}|\mathbf{x}_n]((L_i)^{-1T})_{jk}\right.$$

$$\left. - \sum_{j=1}^{d_z}W_{ij}(L_i)_{jk}\right)^2$$

$$\text{where } \sum_{l=1}^{d_z}(L_i)_{jl}(L_i)_{lk}^T = \sum_{n=1}^{N}2b\lambda(\hat{\xi}_{in})\mathbb{E}[z_{jn}z_{kn}|\mathbf{x}_n]$$

$$\hat{\mu}_i|\hat{\xi}_{in},\hat{W}_{ij} = \frac{1}{\sum_{n=1}^{N}2b\lambda(\hat{\xi}_{in})}\sum_{n=1}^{N}\left(\left(x_{in} - \frac{b}{2}\right) - 2b\lambda(\hat{\xi}_{in})\sum_{j=1}^{d_z}\hat{W}_{ij}\mathbb{E}[z_{jn}|\mathbf{x}_n]\right), \quad (23)$$

where each $(\mathbf{L}_i)$ is the $i^{th}$ (in $d_x$) $(d_z \times d_z)$-Cholesky decomposition. There are $d_x$ total $d_z$-dimensional minimization problems for $\hat{\mathbf{W}}_i$. In these expressions, repeated indices do not imply summations. We have revealed the indices and summations for the sake of clarity. Note that this is an expectation-conditional-maximization algorithm (ECM) due to the structure of successive maximizations (Meng and Rubin 1993). We have suppressed a EM-iteration index in favor of $\hat{\cdot}$ referring to the current iteration's maximum likelihood estimate of the parameter and the absence of a hat $\cdot$ on a parameter referring to the prior iteration's maximum likelihood estimate of the parameter.

The conditionally multinomial distribution Multinomial$(b, \mathbf{f} = \text{softmax}(\mathbf{Wz} + \boldsymbol{\mu}))$ is,

$$p_{b,\mathbf{W},\boldsymbol{\mu}}(\mathbf{x}|\mathbf{z}) = \frac{b!}{\prod_{i=1}^{d_x}x_i!}\exp\left(\mathbf{x}^T(\mathbf{Wz}+\boldsymbol{\mu}) - b\ln\left(\sum_{i=1}^{d_x}\exp(\mathbf{W}_i\mathbf{z}+\mu_i)\right)\right) \quad (24)$$

where $b = \sum_{i=1}^{d_x}x_i$ and $\sum_{i=1}^{d_x}f_i = 1$. To have the appropriate number of parameters for the multinomial distribution, we set the elements of the last row of $\mathbf{W}$ and $\boldsymbol{\mu}$ equal to zero.

Bouchard (2007) introduce the following two-step bound on the function $\ln\left(\sum_{i=1}^{d_x}\exp\left(\eta_i\right)\right)$. This bound builds on the work of Jaakkola and Jordan (1997) and Tipping (1999) and allows for a closed-form approximate EM algorithm for the multinomial factor analysis setting. The bound in Bouchard (2007) is,

$$\ln\left(\sum_{i=1}^{d_x}\exp(\eta_i)\right) \leq \alpha + \sum_{i=1}^{d_x}\ln(1+\exp(\eta_i-\alpha)) = \alpha - \sum_{i=1}^{d_x}\ln\sigma(-\eta_i+\alpha) \quad (25)$$

Next, the variational approximation (Eqn. 19) is applied separately to each $\sigma(-\eta_i+\alpha)$. This is the same as $i$ independent binomial approximations, with one more parameter, $\alpha$, for the overall constraint. The approximate conditional multivariate distribution is,

$$\begin{aligned}
\tilde{p}_{b,\mathbf{W},\boldsymbol{\mu},\boldsymbol{\xi}}(\mathbf{x}|\mathbf{z}) &= \frac{b!}{\prod_{i=1}^{d_x}x_i!}\exp\left(-b\alpha\right)\sigma^b(\xi_i)\exp\left(\mathbf{x}^T\left(\mathbf{W}\mathbf{z}+\boldsymbol{\mu}\right)\right)\\
&\quad \prod_{i=1}^{d_x}\exp\left(-\frac{b}{2}(\mathbf{W}_i\mathbf{z}+\mu_i-\alpha+\xi_i)-b\lambda(\xi_i)((\mathbf{W}_i\mathbf{z}+\mu_i-\alpha)^2-\xi_i^2)\right)
\end{aligned} \quad (26)$$

For $N$ individuals, $\boldsymbol{\alpha}$ is a $1\times N$ vector, and, as in the binomial case, there are $d_x\times N$ variational parameters $\boldsymbol{\xi}$.

Like the binomial case, this approximation provides a lower bound: $\tilde{p}_{b,\mathbf{W},\boldsymbol{\mu},\boldsymbol{\xi}}(\mathbf{x}|\mathbf{z}) \leq p_{b,\mathbf{W},\boldsymbol{\mu}}(\mathbf{x}|\mathbf{z})$ and $\tilde{p}_{b,\mathbf{W},\boldsymbol{\mu},\boldsymbol{\xi}}(\mathbf{x}) \leq p_{b,\mathbf{W},\boldsymbol{\mu}}(\mathbf{x})$. The variational approximation is again quadratic in $\mathbf{z}$, and the EM updates for the approximate complete log-likelihood are analytic, and the conditional probability of $\tilde{p}(\mathbf{z}|\mathbf{x})$ becomes Gaussian.

With this approximation, for individuals $n\in\{1,\ldots,N\}$, the E-steps are,

$$\mathbb{E}[z_{jn}|\mathbf{x}_n] = \sum_{k=1}^{d_z}(C_n)_{jk}\left(\sum_{i=1}^{d_x}W_{ik}\left(x_{in}-\frac{b}{2}-2b\lambda(\xi_{in})(\mu_{in}-\alpha_n)\right)\right) \quad (27)$$

$$\mathbb{E}[z_{jn}z_{kn}|\mathbf{x}_n] = (C_n)_{jk}+\mathbb{E}[z_{jn}|\mathbf{x}_n]\mathbb{E}[z_{kn}|\mathbf{x}_n]$$

$$\text{where } ((C_n)^{-1})_{jk} = \left(\delta_{jk}+2b\sum_{i=1}^{d_x}\lambda(\xi_{in})W_{ij}W_{ik}\right) \quad (28)$$

In these expressions, repeated indices do not imply summations. $\mu_{in}$ is the $(in)^{th}$ element of the matrix of $N$ columns of the $d_x\times 1$ vector $\boldsymbol{\mu}$. $\delta_{jk}$ is the Kronecker delta. Each $\mathbf{C}_n$ is a $d_z\times d_z$ matrix.

The M-steps are,

$$\begin{aligned}
\hat{\xi}_{in}^2|W_{ij},\mu_{in},\alpha_n &= \sum_{k=1}^{d_z}\sum_{j=1}^{d_z}W_{ij}W_{ik}\mathbb{E}[z_{jn}z_{kn}|\mathbf{x}_n]+2\sum_{j=1}^{d_z}W_{ij}\mathbb{E}[z_{jn}|\mathbf{x}_n]\mu_{in}+\mu_{in}^2\\
&\quad -2\alpha_n\sum_{j=1}^{d_z}W_{ij}\mathbb{E}[z_{jn}|\mathbf{x}_n]+\alpha_n^2-2\alpha_n\mu_{in}
\end{aligned}$$

$$\hat{\alpha}_n|\hat{\xi}_{in}, W_{ij}, \mu_{in} = \frac{1}{\sum_{i=1}^{d_x} \lambda(\hat{\xi}_{in})} \left( \sum_{i=1}^{d_x} \lambda(\hat{\xi}_{in}) \left( \sum_{j=1}^{d_z} W_{ij} \mathbb{E}[z_{jn}|\mathbf{x}_n] + \mu_{in} \right) - \frac{1 - \frac{d_x}{2}}{2} \right)$$

$$\hat{W}_{ik}|\hat{\xi}_{in}, \hat{\alpha}_n, \mu_{in} = \min_{W_{ik}} \frac{1}{2} \left( \sum_{j=1}^{d_z} \sum_{n=1}^{N} \left( x_{in} - \frac{b}{2} - 2b\lambda(\hat{\xi}_{in})(\mu_{in} - \hat{\alpha}_n) \right) \mathbb{E}[z_{jn}|\mathbf{x}_n]((L_i)^{-1T})_{jk} \right.$$
$$\left. - \sum_{j=1}^{d_z} W_{ij}(L_i)_{jk} \right)^2$$

$$\text{where } \sum_{l=1}^{d_z} (L_i)_{jl}(L_i)_{lk}^T = \sum_{n=1}^{N} 2b\lambda(\hat{\xi}_{in}) \mathbb{E}[z_{jn} z_{kn}|\mathbf{x}_n]$$

$$\hat{\mu}_i|\hat{\xi}_{in}, \hat{W}_{ij}, \hat{\alpha}_n = \frac{1}{\sum_{n=1}^{N} 2b\lambda(\hat{\xi}_{in})} \sum_{n=1}^{N} \left( \left( x_{in} - \frac{b}{2} \right) - 2b\lambda(\hat{\xi}_{in}) \left( \sum_{j=1}^{d_z} \hat{W}_{ij} \mathbb{E}[z_{jn}|\mathbf{x}_n] - \hat{\alpha}_n \right) \right) \quad (29)$$

where the conventions are the same as for the conditionally binomial case. We choose an ECM approach for **W** and **μ** for the same reasons as for the conditionally binomial case. We also choose an ECM approach for the variational parameters to avoid large matrix inversions.

### 3.3.2 Diverse conditional distributions

For diverse conditional distributions such as (Eqn. 13), with the variational approximations in place for the conditionally binomial covariates and conditionally multinomial covariates, the conditional probability of $\tilde{p}(\mathbf{z}|\mathbf{x})$ remains Gaussian. The E-step for individuals $n \in \{1, \ldots, N\}$ are,

$$\mathbb{E}[z_{jn}|\mathbf{x}_n] = \sum_{k=1}^{d_z} (C_n)_{jk} \sum_{d=1}^{D} \left( \mathbf{I}(\text{type}(d) = \text{normal}) \sum_{i=1}^{d_x^{(d)}} W_{ik}^{(d)} (\Psi^{(d)})_{ii}^{-1} (x_{in}^{(d)} - \mu_{in}^{(d)}) \right.$$
$$+ \mathbf{I}(\text{type}(d) = \text{binomial}) \sum_{i=1}^{d_x^{(d)}} W_{ik}^{(d)} \left( x_{in}^{(d)} - \frac{b^{(d)}}{2} - 2b^{(d)} \lambda(\xi_{in}^{(d)}) \mu_{in}^{(d)} \right)$$
$$\left. + \mathbf{I}(\text{type}(d) = \text{multinomial}) \sum_{i=1}^{d_x^{(d)}} W_{ik}^{(d)} \left( x_{in}^{(d)} - \frac{b^{(d)}}{2} - 2b^{(d)} \lambda(\xi_{in}^{(d)})(\mu_{in}^{(d)} - \alpha_n^{(d)}) \right) \right)$$

$$\mathbb{E}[z_{jn} z_{kn}|\mathbf{x}_n] = (C_n)_{jk} + \mathbb{E}[z_{jn}|\mathbf{x}_n]\mathbb{E}[z_{kn}|\mathbf{x}_n]$$

$$\text{where } ((C_n)^{-1})_{jk} = \left( \delta_{jk} + \sum_{d=1}^{D} \left( \mathbf{I}(\text{type}(d) = \text{normal}) \sum_{i=1}^{d_x^{(d)}} W_{ij}^{(d)} (\Psi^{(d)})_{ii}^{-1} W_{ik}^{(d)} \right.\right.$$
$$\left.\left. + \mathbf{I}(\text{type}(d) \neq \text{normal}) 2b^{(d)} \sum_{i=1}^{d_x^{(d)}} \lambda(\xi_{in}^{(d)}) W_{ij}^{(d)} W_{ik}^{(d)} \right) \right). \quad (30)$$

The M-steps for each $(d)$ datatype's parameters are simply the M-steps for the relevant datatype given in (eqns. 17, 23, 29).

### 3.3.3 Initialization

For each datatype ($d$), we take a warm start initialization of the EM algorithm. In the following, we will omit the datatype index. If $d_x > d_z$, we initialize FA at the probabilistic PCA solution (Tipping 1999),

$$\begin{aligned}
\boldsymbol{\mu}^{(0)} &= \frac{1}{N}\sum_{n=1}^{N} \mathbf{x}_n \\
\mathbf{W}^{(0)} &= \mathbf{U}_{d_z}(\boldsymbol{\Lambda}^2_{d_z} - \sigma^2 \mathbb{1}) \\
\boldsymbol{\Psi}^{(0)} &= \sigma^2 \mathbb{1}
\end{aligned}$$

where $\mathbf{X} - \boldsymbol{\mu}^{(0)} = \mathbf{U}\boldsymbol{\Lambda}\mathbf{V}^T$ is the singular value decomposition

and $\sigma^2 = \dfrac{1}{N(d_x - d_z)} \sum_{i=d_z+1}^{d_x} \Lambda_{ii}^2$ \hfill (31)

and $\mathbf{U}_{d_z}$ is the first $d_z$ columns of $\mathbf{U}$, and likewise for $\boldsymbol{\Lambda}$. If $d_x \leq d_z$, then we initialize $\mathbf{W}^{(0)}$ as a matrix of ones, and $\boldsymbol{\Psi}^{(0)} = (\Lambda_{ii}^2 \mathbf{I}(i = d_x)/N)\mathbb{1}$.

We initialize all variational parameters to one ($\alpha_n = 1, \xi_{in} = 1$), and enforce the appropriate constraints on $\mathbf{W}, \boldsymbol{\mu}$ for the conditionally multinomial distributions.

## 4 Joint factor analysis and exponential Cox proportional hazards model with informative censoring (FA-ECPH-C)

We take the following latent variable model for our observed covariates $\mathbf{x}^{(d)}$ from $D$ different datatypes and survival time with informative censoring. We have a latent variable of dimension $d_z \times 1$ that is normally distributed,

$$\mathbf{z} \sim \mathcal{N}(\mathbb{0}, \mathbb{1}). \tag{32}$$

As in FA (Eqn. 12), the conditional distribution of $\mathbf{x}^{(d)}$ is conditionally normal, binomial, or multinomial.

We take the exponential Cox proportional hazards distribution (Eqn. 2) for our survival time and censoring time distributions. However, in our setting, the covariates are the latent variables $\mathbf{z}$,

$$\begin{aligned}
t|\mathbf{z} &\sim \text{Exp}(\rho_T = \exp\left(\mathbf{w}_T^T \tilde{\mathbf{z}}\right)) \\
c|\mathbf{z} &\sim \text{Exp}(\rho_C = \exp\left(\mathbf{w}_C^T \tilde{\mathbf{z}}\right)),
\end{aligned} \tag{33}$$

where $\tilde{\mathbf{z}}$ is defined as in (Eqn. 1), and $\rho_{T,C}$ refers to the parameters of the exponential distributions.

The model for data generation is,

$$p_\Theta(t, c, \mathbf{x}^{(1)}, \ldots, \mathbf{x}^{(D)}, \mathbf{z}) = p_{\mathbf{w}_T}(t|\mathbf{z}) p_{\mathbf{w}_C}(c|\mathbf{z}) \left( \prod_{d=1}^{D} p_{\Theta^{(d)}}(\mathbf{x}^{(d)}|\mathbf{z}) \right) p(\mathbf{z}) \qquad (34)$$

The key assumption for this model is that, conditioned on the latent variable $\mathbf{z}$, the observed covariates $\mathbf{x}^{(d)}$, $t$, and $c$ are all mutually independent. Note that while $t \perp c|\mathbf{z}$, for generic $\mathbf{w}_C$, $t \not\perp c|\mathbf{x}^{(1)}, \ldots, \mathbf{x}^{(D)}$, and the CAR assumption does not hold. The CAR assumption does hold when $\boldsymbol{\beta}_C = \mathbb{0}$ (see Eqn. 1 for the relation between $\mathbf{w}_C$ and $\boldsymbol{\beta}_C$).

As in Sec. 2, we do not observe $t$ and $c$ directly. Instead we observe $\tilde{t} = \min(t, c)$ and $\delta = \mathbf{I}(\tilde{t} = t)$, where $\mathbf{I}$ is the indicator function. As before, the conditional time-to-event and censoring distribution $p(\tilde{t}, \delta|\mathbf{z})$ for an individual is simply the cumulative probability for each event,

$$p_{\mathbf{w}_T, \mathbf{w}_C}(\tilde{t}, \delta|\mathbf{z}) = \left( p_{\mathbf{w}_T}(\tilde{t}|\mathbf{z}) P_{\mathbf{w}_C}(c > \tilde{t}|\mathbf{z}) \right)^\delta \left( P_{\mathbf{w}_T}(t > \tilde{t}|\mathbf{z}) p_{\mathbf{w}_C}(\tilde{t}|\mathbf{z}) \right)^{1-\delta} \qquad (35)$$

The joint probability of the observed covariates, the time-to-event, the censoring indicator, and the latent variables is,

$$p_\Theta(\tilde{t}, \delta, \mathbf{x}^{(1)}, \ldots, \mathbf{x}^{(D)}, \mathbf{z}) = p_{\mathbf{w}_T, \mathbf{w}_C}(\tilde{t}, \delta|\mathbf{z}) \left( \prod_{d=1}^{D} p_{\Theta^{(d)}}(\mathbf{x}^{(d)}|\mathbf{z}) \right) p(\mathbf{z}), \qquad (36)$$

where we have collected all of the parameters for the $d^{th}$ conditional distribution into $\Theta^{(d)}$ and all of the parameters of the joint distribution into $\Theta = \{\mathbf{w}_T, \mathbf{w}_C, \Theta^{(1)}, \ldots, \Theta^{(D)}\}$.

### 4.1 Inference

With the addition of time-to-event and censoring data, the marginal probability of the data $p_\theta(\tilde{t}, \delta, \mathbf{x}^{(1)}, \ldots, \mathbf{x}^{(D)})$ is not analytic, even with the variational approximations. As a result, the conditional expectations of the latent variables given the data are also not analytic. We calculate the necessary conditional expectations of functions $f(\mathbf{z})$ given $\tilde{t}, \delta, \mathbf{x}^{(1)}, \ldots, \mathbf{x}^{(D)}$ by sampling from the conditional distribution using the Metropolis-Hastings (MH) algorithm (Metropolis *et al.* 1953; Hastings 1970). The use of Monte-Carlo methods to approximate the E-step in the EM algorithm was first introduced by Wei and Tanner (1990). The proposal density we use is $\mathbf{z}'_n|\mathbf{z}_n^{(s)} \sim \mathcal{N}(\mathbf{z}_n^{(s)}, \kappa C_n)$, where $\mathbf{z}'_n$ is the proposal sample for step $(s+1)$ and individual $n$, $C_n$ is the covariance of individual $n$ under the FA-only model (Eqn. 30), and $\kappa$ is a scale parameter. We initialize at $\mathbf{z}_n^{(0)} = \mathbb{E}[\mathbf{z}_n|\mathbf{x}_n]$, the conditional expectation of individual $n$ under the FA-only model (Eqn. 30).

We discard the first $s = \{1, \ldots, 300\}$ burn-in samples and collect the $s = \{301, \ldots, 600\}$ samples for each individual $n$. We use these 300 samples for each individual $n$ to calculate the empirical conditional expectations $\mathbb{E}[f(\mathbf{z})|\tilde{t}, \delta, \mathbf{x}^{(1)}, \ldots, \mathbf{x}^{(D)}]$.

We monitor the acceptance rate $r_a$, effective sample size $n_{eff}$, and convergence parameter $\hat{R}(|\mathbb{E}[\mathbf{z}_n|\mathbf{x}_n]|^2)$ for two parallel Metropolis-Hastings sampling chains for the first E-step for the first-individual to tune the scale parameter $\kappa$ (Gelman *et al.* 2013). The effective sample size is defined in Gelman *et al.* (2013) (Eqn. 11.4), and the convergence parameter $\hat{R}$ of a statistic is defined in Gelman *et al.* (2013) between (eqns. 11.3-4). Starting with $\kappa = 6$ and continuing in descending order ( $\kappa \in \{6, 5.5, 5, 4.5, 4, 3.5, 3, 2.5, 2, 1.5, 1, 0.5, 0.25, 0.1\}$), we select the first scale parameter that has an acceptance rate between $0.134 \leq r_a \leq 0.334$, an effective sample size $n_{eff} \geq 10$, and a convergence parameter $\hat{R}(|\mathbb{E}[\mathbf{z}_n|\mathbf{x}_n]|^2) \leq 1.2$. Gelman *et al.* (2013) contains a comprehensive discussion of these statistics.

The M-step for each $\mathbf{x}^{(d)}$ datatype's parameters is simply the M-step for the relevant datatype given in (eqns. 17, 23, 29). For the time-to-event and censoring datatype M-steps, we perform a Newton-Raphson step (Wu 1983). The subsequent EM algorithm is a generalized EM (GEM) algorithm (Dempster *et al.* 1977). The M-step is the same for $\mathbf{w}_{T,C}$ with $\delta^{T,C}$. We omit the $T, C$ index below. At step $(s+1)$, the M-step is,

$$\mathbf{w}^{(s+1)} = \mathbf{w}^{(s)} + a \left( \sum_{n=1}^{N} \tilde{t}_n \mathbb{E}_{\Theta^{(s)}} \left[ \tilde{\mathbf{z}}_n \tilde{\mathbf{z}}_n^T \exp\left( \left(\mathbf{w}^{(s)}\right)^T \tilde{\mathbf{z}}_n \right) \Big| \tilde{t}, \delta, \mathbf{x}^{(1)}, \ldots, \mathbf{x}^{(D)} \right] \right)^{-1}$$

$$\sum_{n=1}^{N} \left( \delta_n \mathbb{E}_{\Theta^{(s)}}[\tilde{\mathbf{z}}_n | \tilde{t}, \delta, \mathbf{x}^{(1)}, \ldots, \mathbf{x}^{(D)}] - \tilde{t}_n \mathbb{E}_{\Theta^{(s)}} \left[ \tilde{\mathbf{z}}_n \exp\left( \left(\mathbf{w}^{(s)}\right)^T \tilde{\mathbf{z}}_n \right) \Big| \tilde{t}, \delta, \mathbf{x}^{(1)}, \ldots, \mathbf{x}^{(D)} \right] \right)$$

where $\quad 0 < a \leq 1$.

The parameter $a$ sets the scale of the Newton-Raphson step size, $\tilde{\mathbf{z}}$ is defined as in (Eqn. 1), and the expectations are taken with respect to the step $(s)$ parameters. We set $a = 1$. We initialize at,

$$\mathbf{w}_{T,C}^{(0)} = \begin{pmatrix} \ln\left( \frac{\sum_{n=1}^{N} \delta_n^{T,C}}{\sum_{n=1}^{N} \tilde{t}_n} \right) \\ \mathbb{0} \end{pmatrix}. \tag{37}$$

For the results in this note, we stop the approximate GEM algorithm after 10 iterations. We explored using a larger number of iterations but found no improvement in the simulated or cross-validated expected c-index (Sec. 7).

## 4.2 Prediction

Once we have estimates for the maximum likelihood parameters $\hat{\mathbf{w}}_T, \hat{\mathbf{w}}_C, \hat{\Theta}$, we would like to predict a survival time $\hat{t}$ given $\mathbf{x}^{(1)}, \ldots, \mathbf{x}^{(D)}$ and the maximum likelihood parameters $\hat{\mathbf{w}}_T, \hat{\mathbf{w}}_C, \hat{\Theta}$. With the variational approximations for binomial and multinomial distributions, the conditional expectation of the survival time $t$ is analytic. The result is,

$$\hat{t} = \mathbb{E}_{\hat{\mathbf{w}}_T, \hat{\mathbf{w}}_C, \hat{\Theta}}[t|\mathbf{x}^{(1)}, \ldots, \mathbf{x}^{(D)}] = \frac{1}{\hat{\lambda}_T} \exp\left(\frac{1}{2}\hat{\boldsymbol{\beta}}_T^T \hat{\mathbf{C}} \hat{\boldsymbol{\beta}}_T - \mathbb{E}_{\hat{\mathbf{w}}_T, \hat{\mathbf{w}}_C, \hat{\Theta}}[\mathbf{z}^T|\mathbf{x}^{(1)}, \ldots, \mathbf{x}^{(D)}]\hat{\boldsymbol{\beta}}_T\right) \quad (38)$$

where $\hat{\mathbf{C}}$ and $\mathbb{E}_{\hat{\mathbf{w}}_T, \hat{\mathbf{w}}_C, \hat{\Theta}}[\mathbf{z}^T|\mathbf{x}^{(1)}, \ldots, \mathbf{x}^{(D)}]$ are defined in (Eqn. 30), and $\hat{\boldsymbol{\beta}}_T, \hat{\lambda}_T$ are components of $\hat{\mathbf{w}}_T$ in (Eqn. 1). For these predictions, we take the average variational parameter values $\hat{\alpha} = \frac{1}{N}\sum_{n=1}^N \hat{\alpha}_n$ and $\hat{\xi}_i = \frac{1}{N}\sum_{n=1}^N \hat{\xi}_{in}$, since the estimates for the variational parameters $\hat{\boldsymbol{\alpha}}$ and $\hat{\boldsymbol{\xi}}$ were dependent on the learning-set individual $n$.

## 5  Fast Approximation

We find a fast, decoupled approximation to the fully integrative model FA-ECPH-C. The fast approximation is as follows: (1) inference for factor analysis and estimation of the conditional expectation of the latent variables, and then (2) inference for an exponential Cox proportional hazards model with the conditional expectation of the latent variables as covariates. The decoupled inference algorithms are analytic. This provides a significant speed-up compared to the fully integrative model, which requires Metropolis-Hastings simulations for every individual at every E-step.

We tested the fast approximation on the cross validation stage of all four real datasets and all four simulations. The average difference in c-index on the validation sets (integrative - fast approximation) is 0.00, and the root mean-squared error is 0.02. The fully integrative model took approximately 11 minutes on 50 cores to fit and predict for the GBM $0^{th}$ cross-validation set for Model 0, while the fast approximation took around 4 minutes and 1 core. For the LGG $0^{th}$ cross-validation set for Model 0, the fully integrative model took approximately 25 minutes, while the fast approximation took approximately 5 minutes.

The majority of the instances tested had the fast approximation perform equivalently to the the fully integrative model. However, while preparing the simulation studies, we observed some instances where the fully integrative model outperformed the fast approximation. We recommend using the fast approximation for exploratory analysis, and the fully integrative model for final analysis.

# 6 Concordance Index

The concordance index (c-index) was introduced by Harrell *et al.* (1982) as a non-parametric measure of survival time prediction accuracy. In the absence of censoring, the c-index is related to the Wilcoxon-Mann-Whitney U-statistic. We use the generalization of the c-index to account for ties introduced by Kang *et al.* (2015). With the following definition,

$$\tau_{(\tilde{t},\delta),(\hat{t},\hat{\delta})} = \frac{1}{N(N-1)} \sum_{n=1}^{N} \sum_{\substack{m=1 \\ m \neq n}}^{N} \left( \mathbf{I}(\tilde{t}_n \geq \tilde{t}_m)\delta_m - \mathbf{I}(\tilde{t}_n \leq \tilde{t}_m)\delta_n \right) \left( \mathbf{I}(\hat{t}_n \geq \hat{t}_m)\hat{\delta}_m - \mathbf{I}(\hat{t}_n \leq \hat{t}_m)\hat{\delta}_n \right)$$

the c-index that accounts for ties is,

$$c(\tilde{\mathbf{t}}, \boldsymbol{\delta}, \hat{\mathbf{t}}, \hat{\boldsymbol{\delta}} = \mathbb{1}) = \frac{1}{2} \left( \frac{\tau_{(\tilde{t},\delta),(\hat{t},\mathbb{1})}}{\tau_{(\tilde{t},\delta),(\tilde{t},\delta)}} + 1 \right). \tag{39}$$

This reduces to the standard c-index when there are no ties.

# 7 Cross-validation Strategy

The following outlines our cross-validation strategy:

1. Reserve 25% of the individuals (uniformly at random) as a test set. These individuals are not used in cross-validation. Divide the remaining 75% individuals (uniformly at random) into $n_{cv}$ cross-validation sets. Our results use $n_{cv} = 5$.

2. Select a learning model $\hat{M}$. This could be various $d_z$ for FA-ECPH-C or various $L_1$ penalties for ECPH-$L_1$.

3. For each $n_{cv}$ cross-validation set $v$, the learning set $l$ is the remaining $(n_{cv} - 1)/n_{cv}$ cross-validation sets. For each set $v$:

   (a) Learn the parameters $\hat{\Theta}_l$ for learning model $\hat{M}$ from the learning set $l$. In this section, $\Theta$ is a collection of all the model parameters. If $\hat{M}$ is a FA-ECPH-C model, use the approximate GEM algorithm outlined in Sec. 4.1. If $\hat{M}$ is a ECPH-$L_1$ model, use the iterative least-squares procedure outlined in Sec. 2.1.

   (b) Predict $\hat{t}_v = \mathbb{E}_{\hat{\Theta}_l}[t|\mathbf{x}_v^{(1)}, \ldots, \mathbf{x}_v^{(D)}]$ for all $N$ individuals in the validation set $v$ (for $\hat{M} \in$ FA-ECPH-C, Eqn. 38 or Eqn. 6 for $\hat{M} \in$ ECPH-C-$L_1$).

   (c) Calculate the c-index on the validation set given the learning set,
   $c(v)|l = c(\mathbf{t}_v, \mathbb{1}, \hat{\mathbf{t}}_v, \mathbb{1})|\hat{\Theta}_l$ (Eqn. 39).

4. Calculate the mean and standard deviation for the cross-validated c-index.

# 8 Model Selection Strategy

We use the following conservative selection criteria for "best" c-index:

- The model with largest mean c-index, as long as no other model's standard deviation is contained within the standard deviation of the largest mean c-index.

- The model with the next largest mean c-index with a standard deviation contained within the standard deviation of the of the largest mean c-index.

- And so on, for additional nesting of standard deviations.

# 9 Additional Figures

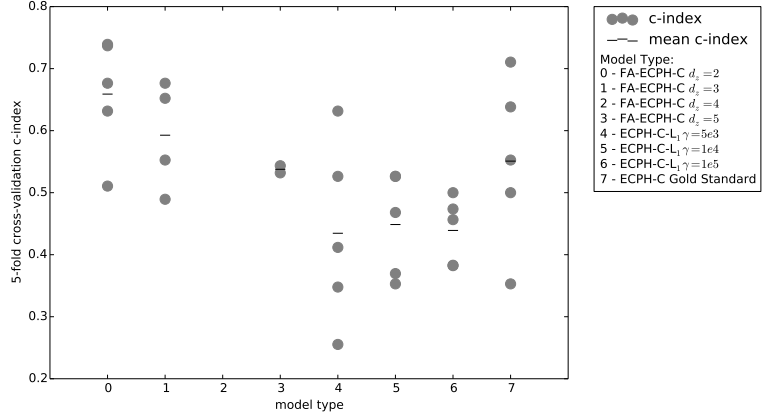

Figure 1: Results for the GBM 5-fold cross-validation latent dimension search for FA-ECPH-C, comparison to ECPH-C-$L_1$, and gold standard ECPH-C. Model types are as follows. Models $0-3$ are FA-ECPH-C and have, in order, $d_z = \{2, 3, 4, 5\}$. Models $1, 2, 3$ are eliminated from the model selection search because on at least 1 out of the 5 cross-validation groups, the EM algorithm approaches a Heywood case. Models $4-6$ are ECPH-C-$L_1$ and have, in order, $\gamma = \{5e3, 1e4, 1e5\}$, which selects on average $\{14, 9, 2\}$ relevant covariates. Model 7 is the gold standard ECPH-C model.

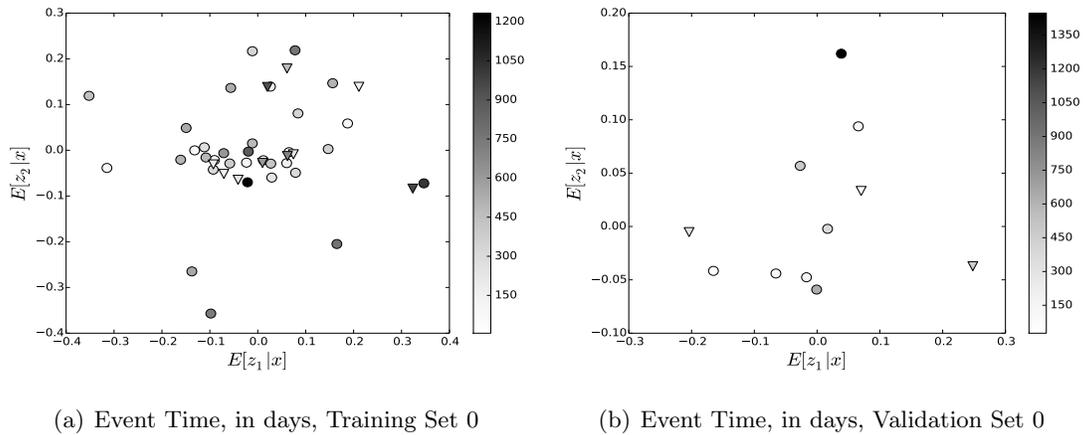

(a) Event Time, in days, Training Set 0

(b) Event Time, in days, Validation Set 0

Figure 2: GBM latent projections $\mathbb{E}[\mathbf{z}|\mathbf{x}^{(1)}, \ldots, \mathbf{x}^{(D)}]$ of the $0^{th}$ cross-validation training and validation cohort for the FA-ECPH-C $d_z = 2$ model. Circles represent uncensored observations, and triangles represent censored observations.

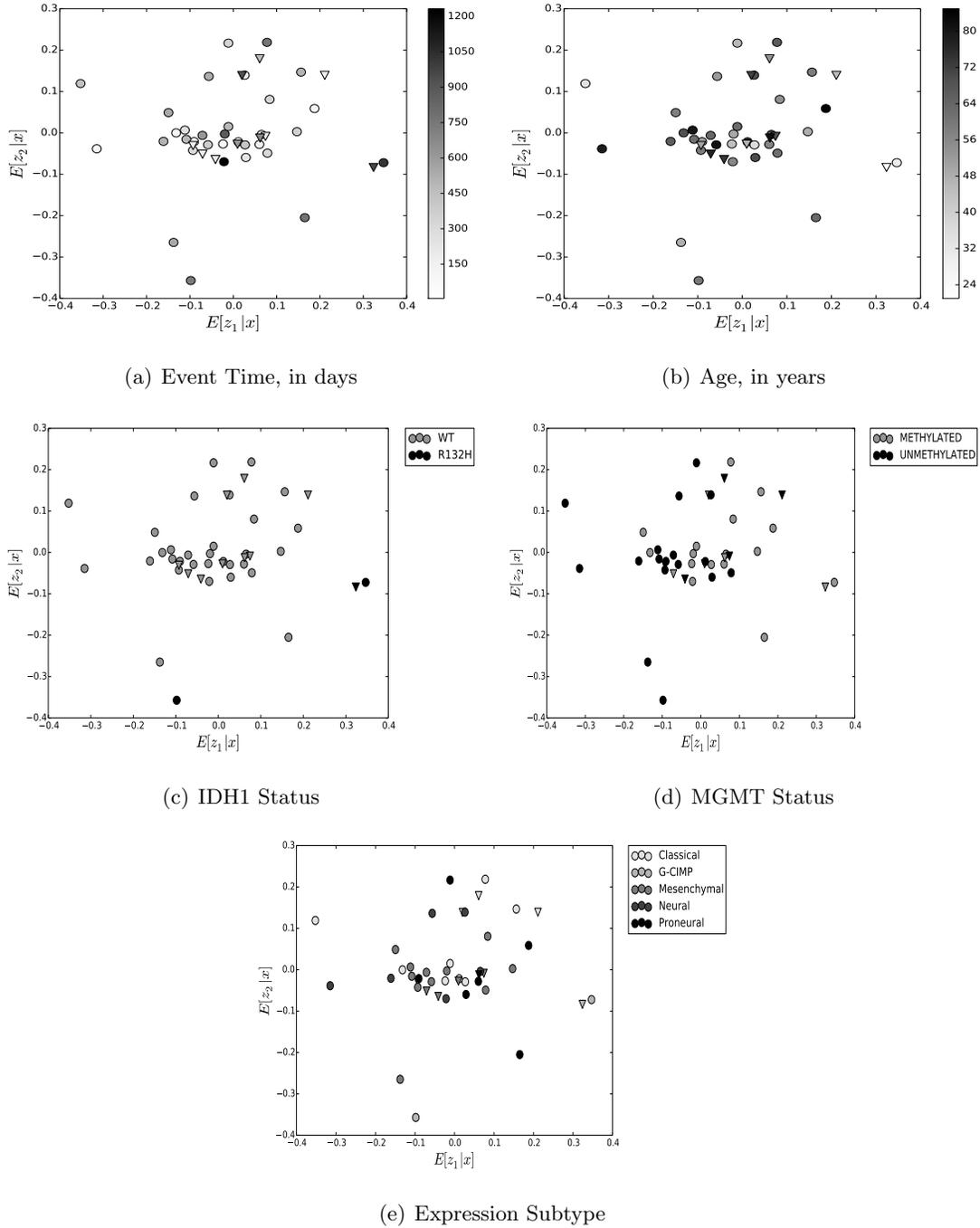

(a) Event Time, in days

(b) Age, in years

(c) IDH1 Status

(d) MGMT Status

(e) Expression Subtype

Figure 3: GBM latent projections $\mathbb{E}[\mathbf{z}|\mathbf{x}^{(1)}, \ldots, \mathbf{x}^{(D)}]$ of the $0^{th}$ cross-validation training cohort for the FA-ECPH-C $d_z = 2$ model. Circles represent uncensored observations, and triangles represent censored observations.

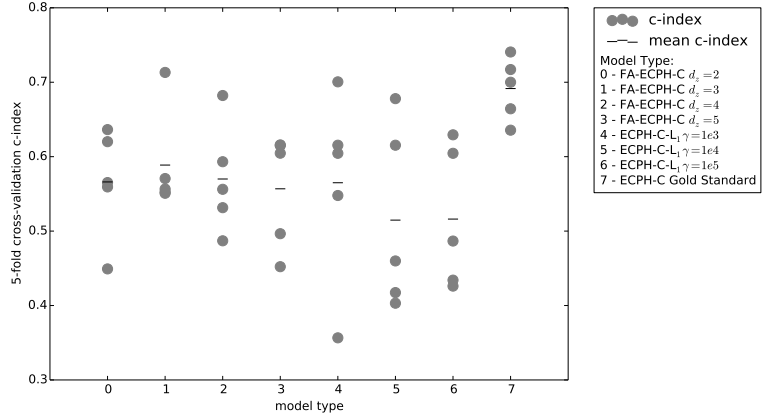

Figure 4: Results for the LUAD 5-fold cross-validation latent dimension search for FA-ECPH-C, comparison to ECPH-C-$L_1$, and gold standard ECPH-C. Model types are as follows. Models $0-4$ are FA-ECPH-C and have, in order, $d_z = \{2, 3, 4, 5\}$. Models $4-6$ are ECPH-C-$L_1$ and have, in order, $\gamma = \{1e3, 1e4, 1e5\}$, which selects on average $\{39, 12, 3\}$ relevant covariates. Model 7 is the gold standard ECPH-C model.

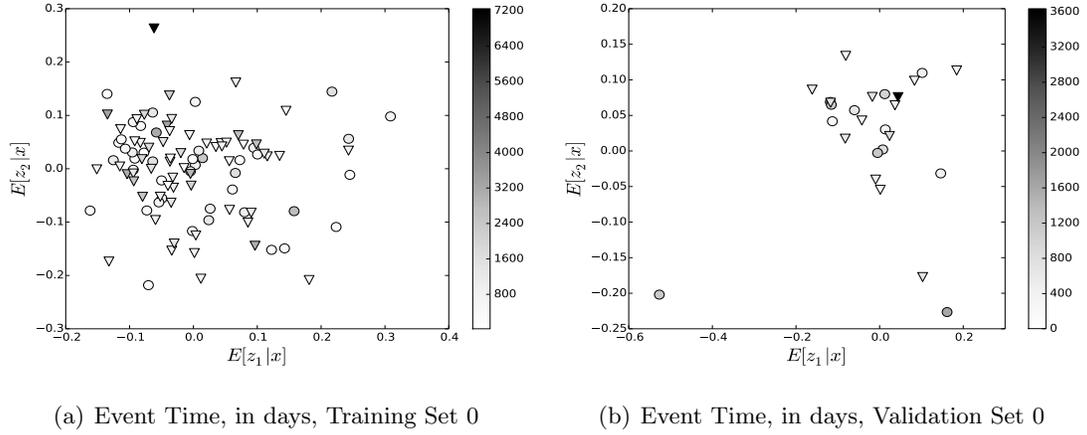

(a) Event Time, in days, Training Set 0

(b) Event Time, in days, Validation Set 0

Figure 5: LUAD latent projections $\mathbb{E}[\mathbf{z}|\mathbf{x}^{(1)}, \ldots, \mathbf{x}^{(D)}]$ of the $0^{th}$ cross-validation training and validation cohort for the FA-ECPH-C $d_z = 2$ model. Circles represent uncensored observations, and triangles represent censored observations.

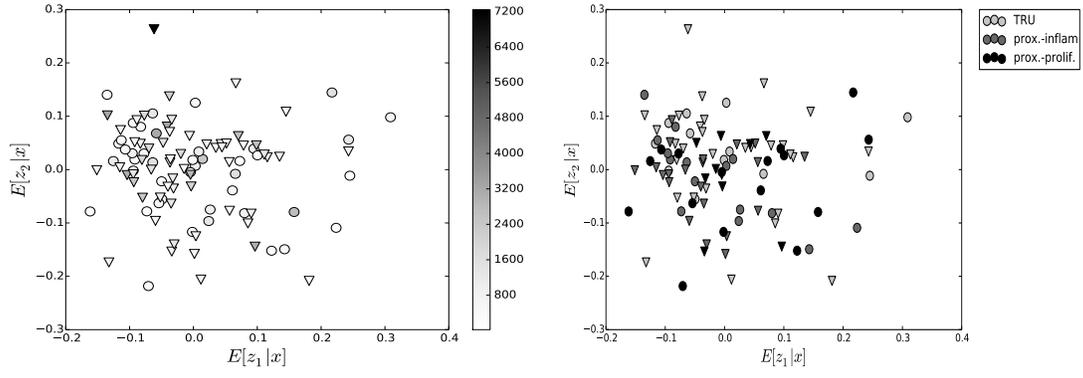

(a) Event Time, in days

(b) Expression Subtype

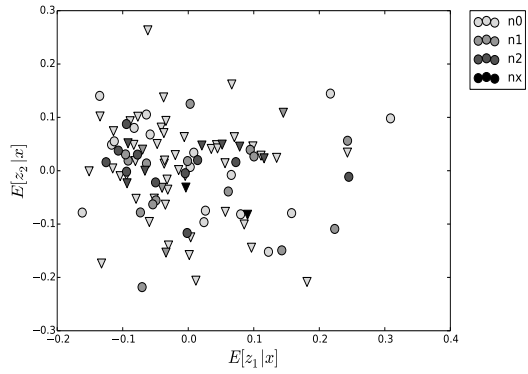

(c) Pathology N-stage

Figure 6: LUAD latent projections $\mathbb{E}[\mathbf{z}|\mathbf{x}^{(1)},\ldots,\mathbf{x}^{(D)}]$ of the $0^{th}$ cross-validation training cohort for the FA-ECPH-C $d_z = 2$ model. Circles represent uncensored observations, and triangles represent censored observations.

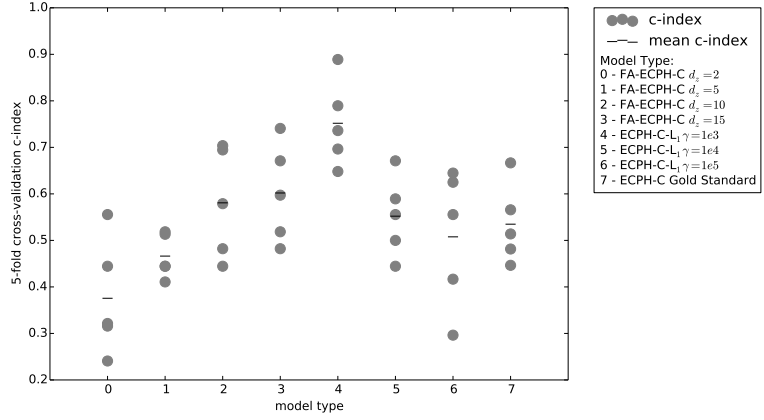

Figure 7: Results for the LUSC 5-fold cross-validation latent dimension search for FA-ECPH-C, comparison to ECPH-C-$L_1$, and gold standard ECPH-C. Model types are as follows. Models $0-3$ are FA-ECPH-C and have, in order, $d_z = \{2, 5, 10, 15\}$. Models $4-6$ are ECPH-C-$L_1$ and have, in order, $\gamma = \{1e3, 1e4, 1e5\}$, which selects an average of $\{29, 11, 6\}$ relevant covariates. Model 7 is the gold standard ECPH-C model.

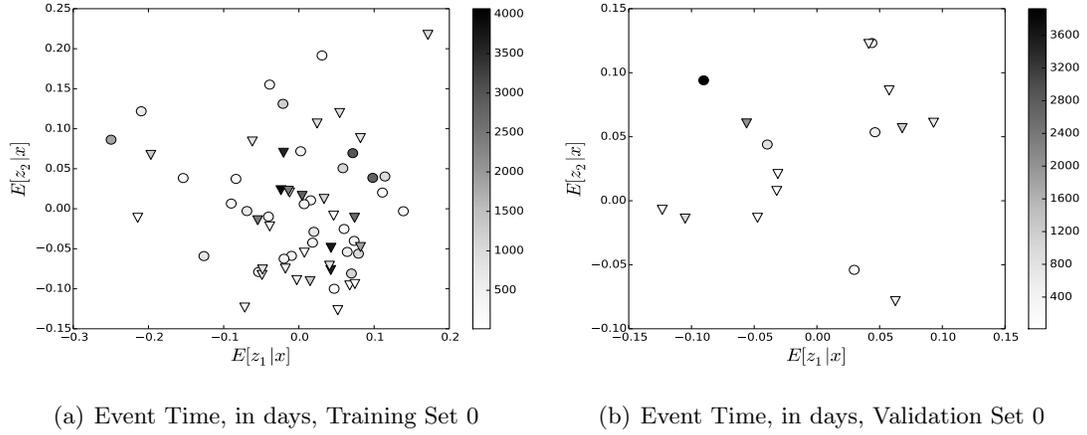

(a) Event Time, in days, Training Set 0

(b) Event Time, in days, Validation Set 0

Figure 8: LUSC latent projections $\mathbb{E}[\mathbf{z}|\mathbf{x}^{(1)}, \ldots, \mathbf{x}^{(D)}]$ of the $0^{th}$ cross-validation training and validation cohort for the FA-ECPH-C $d_z = 2$ model. Circles represent uncensored observations, and triangles represent censored observations.

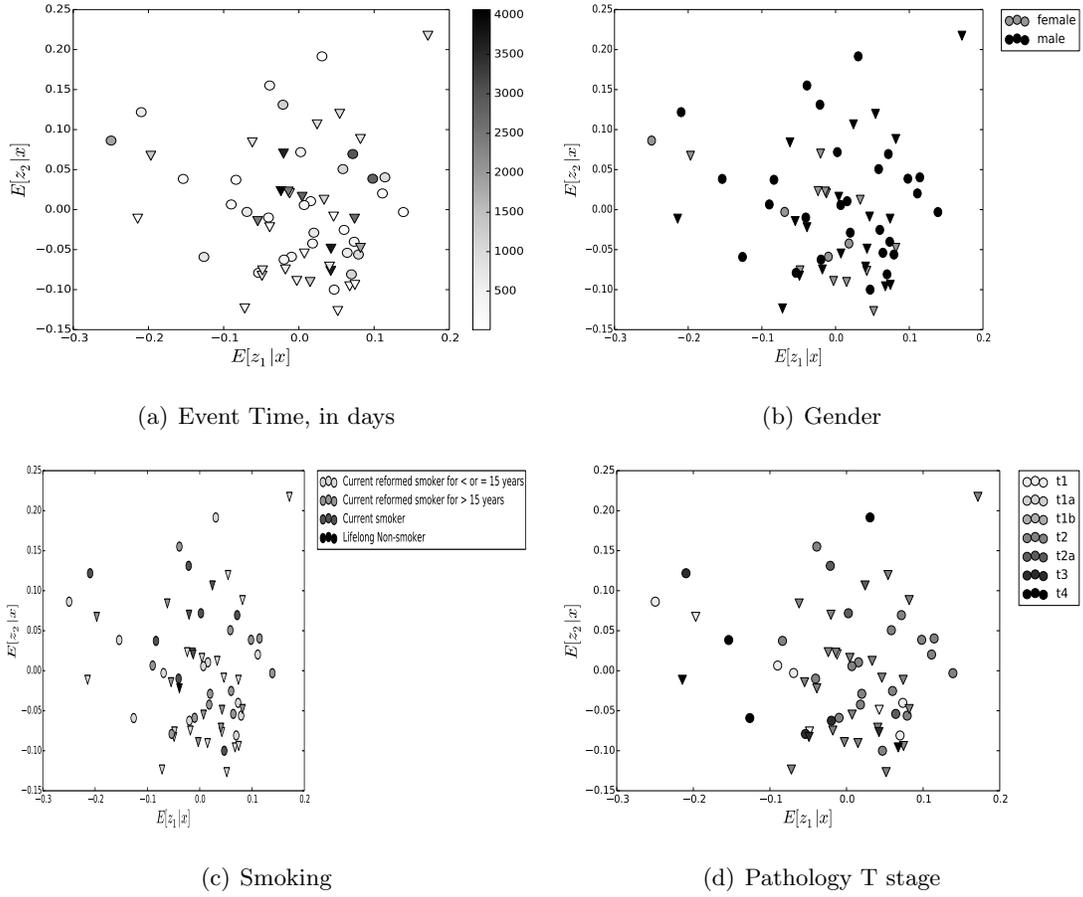

Figure 9: LUSC latent projections $\mathbb{E}[\mathbf{z}|\mathbf{x}^{(1)}, \ldots, \mathbf{x}^{(D)}]$ of the $0^{th}$ cross-validation training cohort for the FA-ECPH-C $d_z = 2$ model. Circles represent uncensored observations, and triangles represent censored observations.